\documentclass[pra,twocolumn,nofootinbib,superscriptaddress]{revtex4}
\usepackage{amssymb,amsmath}
\usepackage{mathptmx}
\usepackage[pdftex]{graphicx}
\usepackage[usenames,dvipsnames]{color}
\usepackage[colorlinks=true, citecolor=blue, linkcolor=RubineRed, urlcolor=Emerald]{hyperref}

\begin{document}
\newcommand{\be}{\begin{equation}}
\newcommand{\ee}{\end{equation}}
\newcommand{\R}[1]{\textcolor{red}{#1}}

\title{Suppression of quantum-radiation-pressure noise in an optical spring}
\author{W.\ Zach Korth}
\affiliation{LIGO Laboratory, California Institute of Technology, Pasadena, CA  91125}
\author{Haixing Miao}
\affiliation{LIGO Laboratory, California Institute of Technology, Pasadena, CA  91125}
\affiliation{Theoretical Astrophysics 350-17, California Institute of Technology, Pasadena, CA  91125}
\author{Thomas Corbitt}
\address{Department of Physics \& Astronomy, Louisiana State University, Baton Rouge, LA  70803}
\author{Garrett D.\ Cole}
\affiliation{Vienna Center for Quantum Science and Technology (VCQ), Faculty of Physics, University of Vienna, Austria}
\author{Yanbei Chen}
\affiliation{Theoretical Astrophysics 350-17, California Institute of Technology, Pasadena, CA  91125}
\author{Rana X.\ Adhikari}
\affiliation{LIGO Laboratory, California Institute of Technology, Pasadena, CA  91125}

\date{July 30, 2013}

\begin{abstract}

Recent advances in micro- and nanofabrication techniques have led to corresponding improvement in the performance 
of optomechanical systems, which provide a promising avenue towards quantum-limited
metrology and the study of quantum behavior in macroscopic mechanical objects.
One major impediment to reaching the quantum regime is thermal excitation, which
can be overcome for sufficiently high mechanical quality factor $Q$. Here, we propose a
method for increasing the effective $Q$ of a mechanical resonator by stiffening it via the
optical spring effect exhibited by linear optomechanical systems, and show how the associated
quantum radiation pressure noise can be evaded by sensing and feedback control.
In a parameter regime that is attainable with current technology, this method allows for realistic quantum cavity 
optomechanics in a frequency band well below that which has been realized
thus far.

\end{abstract}

\maketitle

\section{Introduction}
Catalyzed by vast improvements in micro- and nanofabrication processes,
the field of cavity optomechanics has seen a recent boom in interest\,\cite{Kippenberg2008,Marquardt:2009fk,Aspelmeyer_review}.
In addition to providing a means for quantum-limited force measurements\,\cite{BK92}, e.g., in
gravitational-wave detection\,\cite{LIGO_review} and scanning probe microscopy, optomechanical devices can also be 
used to probe the quantum behavior of mechanical systems. Recently, several experiments have demonstrated
the cooling of a resonator down to its quantum ground state via cryogenics or optomechanical
interaction\,\cite{Teufel2011,OConnell2010,Chan:2011uq}. In addition, more than one
group~\cite{Weis10122010, Safavi-Naeini:2011kx} has demonstrated the so-called optomechanically
induced transparency (OMIT) effect, an analog of the electromagnetically induced transparency
(EIT)~\cite{Boller:1991yq, Harris:1997ly} effect observed in atomic systems. This effect can be used
to make narrow-band quantum filters, e.g., to effect the frequency-dependent phase rotation of squeezed
light injected to enhance the sensitivity of quantum noise-limited interferometric gravitational
wave detectors\,\cite{Kimble:2001kx, Yiqiu2012}. This also opens up the possibility for the processing
and storing of nonclassical states of light through coherent transfer of quantum states between light
and a mechanical oscillator, a technique that would find much use in the emergent field of quantum
information processing.

The ubiquitous bath of thermal energy presents a major obstacle to these efforts, randomly
exciting a system and masking its underlying quantum nature. A characteristic figure
of merit for quantifying this thermal decoherence effect is given by the ratio of the thermal
occupation number $\bar n_{\rm th}$ and the mechanical quality factor $Q$:
\begin{equation}
\frac{\bar n_{\rm th}}{Q}=\frac{k_B T}{\hbar \omega_m Q}\propto (Q f)^{-1},
\end{equation}
where $k_B$ is the Boltzmann's constant and $f=\omega_m/2\pi$ is the mechanical frequency.
When this ratio becomes smaller than one, the oscillator quantum state will survive longer
than one oscillation period before the thermal effect destroys it.

As is apparent from Eq. (1), the quantum state lifetime is ultimately limited
by the product of the quality factor $Q$ and mechanical frequency $f$. A significant body
of research has focused on increasing this ``$Qf$ product'' for a wide range of mechanical systems. If $Q$ were truly a frequency-independent quantity---as in the ``structural damping'' model as described by Saulson~\cite{Saulson:1990qy}---then moving to higher eigenfrequencies would lead to an immediate improvement. In the opposite direction, there are many experiments that would benefit from the use of low-frequency (sub-kHz) resonators. A number of bulk structures have been found to exhibit extremely high $Q$ in this frequency range~\cite{McGuigan:1978fj,Ageev:2004yq}; unfortunately, such systems tend to have relatively large (gram- to kg-scale) effective masses, making them unsuitable for typical optomechanics experiments. The realization of sub-microgram effective masses requires the use of nanofabricated resonators. In practice, excess damping from surface effects~\cite{Mihailovich:1995fk}, phonon tunneling loss~\cite{Cole:2011rt} or intrinsic mechanisms such as thermoelastic~\cite{Zener:1938zr} and Akhiezer damping~\cite{Akhiezer:1939ly} limits the achievable $Q$ and thus the $Qf$ product in these devices. In addition, we add the further requirement that the desired system exhibit excellent optical quality (i.e., high reflectivity owing to low scatter loss and absorption), which limits the resonator options considerably, especially in light of the fact that typical dielectric materials used to create multi-layer optical coatings (e.g., SiO$_2$/Ta$_2$O$_5$) exhibit low mechanical quality factors~\cite{Harry:2002ve}.
Here, we propose a method for using the optical spring effect in linear optomechanical devices\,\cite{Braginsky1997, Braginsky:1999ys, Buonanno2002, Corbitt:2007fk, Corbitt:2007lr, Di-Virgilio:2006fk} to increase the
effective $Q$ of a given mechanical resonator, while simultaneously suppressing the quantum radiation pressure noise that would normally be imparted by the optical fields. This technique should facilitate the creation
of an oscillator with a $Qf$ product considerably higher than those available today, enabling useful applications in
quantum metrology and also creation of long-lived quantum states at lower frequencies than were previously practical.

The concept in this paper makes use of the fact that when a strong optical spring is linearly coupled to a mechanical resonator,
the resonator's Hamiltonian becomes augmented or even dominated by contributions from the radiation
pressure forces of the optical fields. In this way, the bare resonator's thermal noise is ``diluted'' by the ratio
of the intrinsic elastic energy to that stored in the optical field\,\cite{Corbitt:2007lr}. Typically, the modification of a resonator's dynamics via linear coupling is accompanied by excess noise from quantum
back-action---the quantum fluctuations of the radiation pressure, in our case. This has been identified as
a serious issue in the strong dilution regime by Chang {\it et al.}~\cite{Chang:2012fj} and
Ni {\it et al.}~\cite{Ni:2012qy}, who instead propose to achieve optical dilution by using a nonlinear quadratic optical potential to trap a partially reflective membrane \cite{Thompson:2008fk}, which would be immune to linear quantum back-action. The device we propose evades such parasitic quantum back-action by detecting it in the outgoing field and actively
feeding back to the system, resulting in a nearly noise-free optical spring. Since this method allows
for straightforward coupling of the diluted mechanical resonator to an external optical system from the reverse side
of the resonator, it can be used as a ``black-box'' effective mechanical resonator of exceptionally high quality.

While superficially similar to another intensity feedback scheme described by Buchler {\it et al.}\,\cite{Buchler:1999fk}, the technique described here differs in at least two important ways. For one, the critical coupling of the optical cavity employed in that previous work ensures that only half of the fluctuations responsible for the radiation pressure noise are measured (leading to a maximal ideal suppression factor of 2), while in our case all fluctuations are measured, leading to an unbounded suppression factor in the ideal case. More fundamentally, their technique is not applicable to a detuned optical cavity, and is therefore unsuitable for use with an optical spring.

\section{Optical spring}
\begin{figure}[!t]
\begin{center}
\includegraphics[width=\columnwidth]{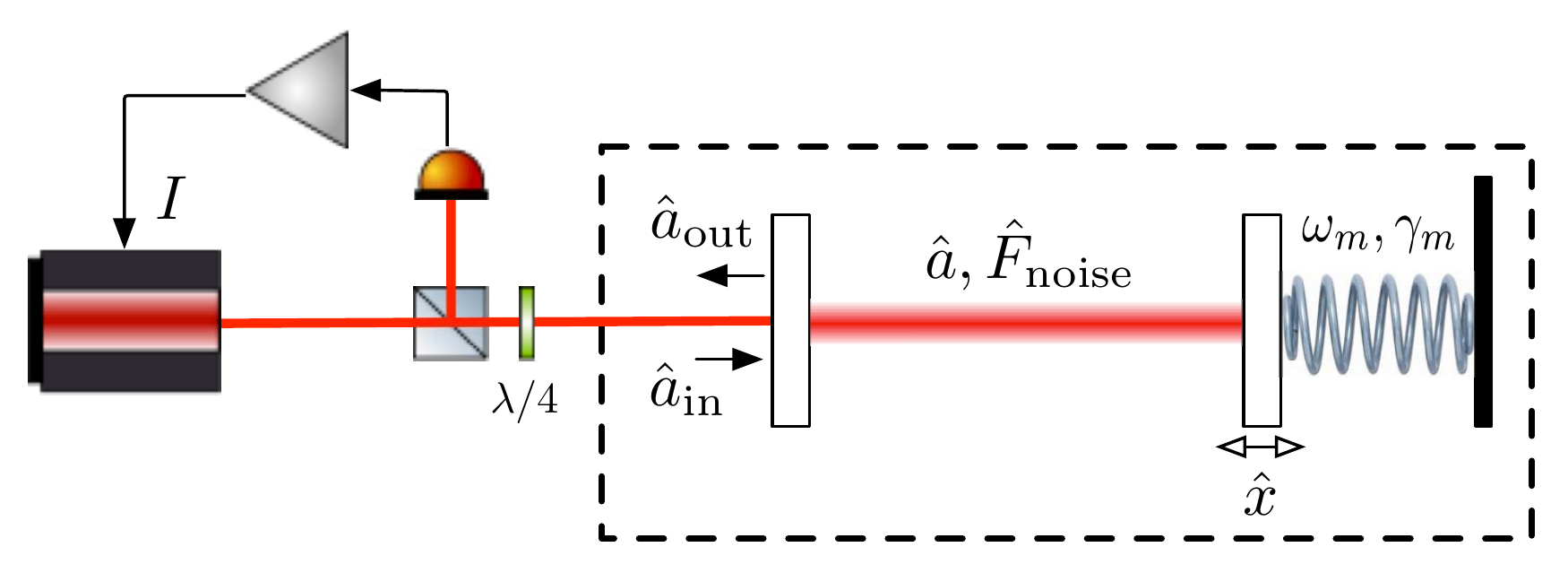}
\caption{Simplified experimental layout, with the canonical optomechanical system shown within the dashed box. Input vacuum fluctuations drive the cavity mode, which
in turn exerts radiation pressure forces on the mechanical resonator forming the cavity boundary.
The output mode of the cavity is sensed with a photodetector, and---in the relevant parameter regime---the measured power contains the radiation pressure fluctuations that drive the resonator with minimal sensitivity to the resonator position. This signal can therefore can be used as an error signal for feeding back to the laser amplitude to suppress the radiation pressure noise on the resonator.}
\label{fig:simple_diagram}
\end{center}
\end{figure}

The canonical optomechanical system is shown in the dashed box in
Fig.\, \ref{fig:simple_diagram}. In such a system, the ``optical spring'' effect arises from
dynamical back-action of the optical cavity field on the mechanical oscillator forming one
cavity boundary. The mechanical oscillator displacement $\hat x$ is coupled to
the cavity field $\hat a$ via radiation pressure, as described by the following
interaction Hamiltonian\,\cite{Marquardt2007}:
\begin{equation}
\hat{\cal H}_{\rm int}=\hbar G_0  \hat x (\bar a^* \hat a + \bar a \hat a^{\dag}) \equiv -\hat x \hat F_{\rm rad}.
\end{equation}
The coupling constant is $G_0= \omega_c/L$; $\bar a$ is the classical
mean amplitude of $\hat a$ due to coherent driving of an external laser; $\omega_c$ is the
cavity resonant frequency; $L$ is the cavity length. When the frequency of the external
laser $\omega_0$ that drives the cavity field is detuned from $\omega_c$, $\hat F_{\rm rad}$
depends on the oscillator displacement, creating a mechanical response that mimics a spring. More specifically, $\hat F_{\rm rad}$ in the
frequency domain can be written as (see Supplemental Material at [URL here] for a more detailed derivation):
\be\label{eq_rad}
\hat F_{\rm rad}(\omega) =- K_{\rm os}(\omega) \hat x(\omega) +\hat F_{\rm noise}(\omega),
\ee
where the optical spring coefficient $K_{\rm os}$ is approximately given by
\be\label{eq_opt}
K_{\rm os}\approx -\frac{2\hbar G_0^2|\bar a|^2 \Delta}{\Delta^2+\gamma^2} - \frac{4i\hbar
G_0^2|\bar a|^2\gamma\Delta \omega}{(\Delta^2+\gamma^2)^2}\equiv m \omega_{\rm os}^2 - i\,m \Gamma_{\rm os}\omega\,,
\ee
with the cavity detuning $\Delta\equiv \omega_c-\omega_0$ and $\gamma$ being the cavity bandwidth.
Here, the approximation is taken for the case of large detuning and cavity bandwidth, which we will show to be the relevant parameter regime for realization of this idea. In addition, we have introduced the optical
spring frequency $\omega_{\rm os}$ and the optical damping $\Gamma_{\rm os}$.
As we can see, when the detuning is negative, i.e., $\Delta <0$, the optical rigidity is real and positive,
and the optical damping is negative $\Gamma_{\rm os}$ (heating), and vice versa. By introducing
an additional driving field with a different detuning frequency, one can create the so-called stable double optical
spring\,\cite{Corbitt:2007fk}, which exhibits both positive rigidity and positive damping (we will elaborate on
this issue later). The optical spring modifies the mechanical susceptibility $\chi_0(\omega)$, defined through
$\chi_0(\omega)\equiv \hat x(\omega)/\hat F(\omega)$, from its original value $
\chi_0^{-1}(\omega)=-m(\omega^2 +i\gamma_m \omega -\omega_m^2)$
to an effective one:
\be\label{eq_reff}
\chi_{\rm eff}^{-1}(\omega)=-m[\omega^2+i(\gamma_m +\Gamma_{\rm os})\omega - (\omega_m^2 +\omega_{\rm os}^2)].
\ee
For a strong optical spring $\omega_{\rm os}\gg \omega_m$, we can significantly stiffen the mechanical
oscillator with the restoring energy from the optical field.

One immediate issue with this approach comes from the quantum radiation pressure noise
$ \hat F_{\rm noise}(\omega)$ in Eq.\,\eqref{eq_rad}, which arises from quantum fluctuation
of the optical field:
\be\label{eq_fnoise}
\hat F_{\rm noise}(\omega) \equiv \frac{ 2\hbar G_0|\bar a|\sqrt{\gamma}}{\sqrt{\gamma^2+\Delta^2}}\left[\frac{(\gamma^2+\Delta^2-i\,\gamma\,\omega)\hat v_1+i\,\Delta\,\omega\, \hat v_2}{(\omega-\Delta +i\gamma)(\omega+\Delta+i\gamma)}\right],
\ee
where $\hat v_1\equiv (\hat a_{\rm in}+\hat a_{\rm in}^{\dag})/\sqrt{2}$ and
$\hat v_2\equiv (\hat a_{\rm in}-\hat a_{\rm in}^{\dag})/\sqrt{2}i$ are the amplitude and phase
quadratures of the input optical field. This additional noise term will increase the effective temperature of the thermal bath, and drive
the mechanical oscillator away from the quantum regime, as pointed out by Chang {\it et al.}~\cite{Chang:2012fj}.
In the large bandwidth and detuning regime, this reduces to
\be\label{eq_BA}
\hat F_{\rm noise} (\omega)\approx-\frac{2\hbar G_0|\bar a|\sqrt{\gamma}}{\sqrt{\gamma^2+\Delta^2}}\hat v_1(\omega),
\ee
indicating that the radiation pressure noise is dominated by fluctuations in the amplitude quadrature of the input field. The strength of this noise can be quantified by its spectral density:
\be\label{eq_SF}
S_F(\omega) \approx \frac{4 \hbar^2 G_0^2 |\bar a|^2 \gamma}{\gamma^2 + \Delta^2}.
\ee
From the above expression and Eq.\,\eqref{eq_opt}, we learn the optical rigidity (real part
of $K_{\rm os}$) scales with the optomechanical coupling strength in the same way as the radiation pressure noise:
\be
K_{\rm os},\,S_{F}\propto G_0^2 |\bar a|^2.
\ee
Essentially, this means that an increase in the optical spring frequency is accompanied by
an increase in the radiation pressure noise when we scale up the optical power.

\section{Evading quantum radiation pressure noise}
To solve the aforementioned issue, we make use of the fact that
the output field emerging from the cavity contains information about the quantum
radiation pressure noise that has been imposed onto the mechanical oscillator. In particular, as we will show, for the large bandwidth and detuning limit, the power fluctuations in the output field originate from the same quadrature that is responsible for the radiation pressure noise.  The photodetector measures these fluctuations, and by feeding this signal back to the mechanical
oscillator with the correct filter, we can therefore evade the quantum radiation pressure noise and achieve a nearly noiseless optical spring. Note that this does not violate the fundamental principle of quantum measurement---any
linear continuous measurement of a dynamical variable that does not commute at different times
(non-conservative) is associated with quantum back-action on that variable\,\cite{BK92}; here, we only
sense the quantum radiation pressure noise and have almost no sensitivity to the mechanical
displacement, and that is why we can evade such back-action noise.

To elaborate on this idea, we use the standard input-output relation for this system $\hat a_{\rm out}(\omega) =-\hat a_{\rm in}(\omega)+\sqrt{2\gamma}\, \hat a(\omega)$, and it, for high bandwidth and detuning, gives [refer to the Appendix for more detail]:
\be\label{eq_ioq}
\hat a_{\rm out}(\omega) \approx -\frac{\Delta + i\gamma}{\Delta - i\gamma} \hat a_{\rm in}(\omega) - \frac{\sqrt{2 \gamma} G_0 \bar a}{\Delta - i\gamma} \hat x(\omega)\,.
\ee
Accompanying these input fluctuations is a classical mean amplitude, $\bar a_{\rm in}$, and we can define a phase reference for the system by setting this field to be real and positive. This field also receives a phase shift upon interaction with the cavity:
\be\label{eq_ioc}
\bar a_{\rm out} = -\bar a_{\rm in} + \sqrt{2 \gamma} \bar a = -\frac{\Delta + i\gamma}{\Delta - i\gamma} \bar a_{\rm in}.
\ee

The power fluctuation measured by a photodetector placed at the cavity output reads: $
\delta \hat P\equiv \bar a_{\rm out}^* \hat a_{\rm out} + \bar a_{\rm out} \hat a_{\rm out}^\dag\,. $
In our stated limit and in the frequency domain, this fluctuating piece is given by
\be
\delta \hat P(\omega) = \bar a_{\rm out}^* \hat a_{\rm out}(\omega) + \bar a_{\rm out} \hat a_{\rm out}^\dag(\omega) \approx \sqrt{2} \bar a_{\rm in} \hat v_1(\omega).
\ee
Therefore, due to the common phase rotation experienced by the DC and fluctuating components  [c.f. Eqs.\,\eqref{eq_ioq} and \eqref{eq_ioc}], the output power is still a measure of the amplitude fluctuations of the input field. As shown in Eq.\,\eqref{eq_BA}, it is this quadrature responsible for the radiation pressure back-action on the resonator, and so the noise can be suppressed by feeding this signal back to the amplitude of the pump laser.

\section{Residual radiation pressure noise}
While strong radiation pressure noise cancelation can be achieved using this technique, a small fraction cannot be canceled owing to two effects: (i) {\it optical loss} due to imperfection of the cavity and non-unity quantum efficiency in photodetection, which will introduce
vacuum noise that is uncorrelated with $\hat v_1$ and $\hat v_2$; (ii) {\it finite cavity bandwidth and detuning} that modifies the input-output relation to give residual parasitic sensitivity to the oscillator displacement $\hat x$, which
we have thus far ignored by assuming very large bandwidth and detuning. In actual experimental setups,
there is always certain amount of optical loss, and the bandwidth and detuning are both finite.

By taking these effects into account, we see that the total measured power fluctuation is
\be
\delta \hat P(\omega) =  \sqrt{2} \bar a_{\rm in} \hat v_1(\omega) +\delta \hat P_{\epsilon}(\omega) + \delta \hat P_{\eta}(\omega)+ \delta \hat P_x(\omega)\,.
\ee
Here, the second term
\be
\delta \hat P_{\epsilon}(\omega) =  \frac{2\sqrt{2\gamma \gamma_\epsilon} \bar a_{\rm in}}{\gamma^2 + \Delta^2} (\gamma \hat v_1' - \Delta \hat v_2') \,
\ee
arises from the vacuum fluctuation $\hat v'_{1,2}$ (uncorrelated with $\hat v_{1,2}$) due to optical loss in the cavity, and  $\gamma_{\epsilon}\equiv c\,\epsilon/(4L)$ with $\epsilon$ being the round-trip optical loss in the cavity; the third term, 
\be
\delta \hat P_{\eta}(\omega) \approx \sqrt{2} \bar a_{\rm in}\sqrt{1-\eta}\,\hat n,
\ee
comes from the non-unity quantum efficiency, $\eta$, of the photodetector (here, $\hat n$ is the vacuum fluctuation associated with this loss port); the last term represents the parasitic position sensitivity:
\be\label{eq_dIx}
\delta \hat P_x(\omega) = - \frac{2 G_0 |\bar a|^2 \Delta (2\gamma_\epsilon - i\omega)}{\gamma^2 + \Delta^2} \hat x(\omega)\,,
\ee
which arises both from the intracavity loss and from the first-order correction to the frequency dependence due to finite bandwidth and detuning, and the associated quantum back-action (radiation pressure) noise reads:
\be
\hat F_{\delta \hat P_x}=-\frac{2\hbar G_0\sqrt{\gamma}|\bar a|}{\sqrt{\gamma^2+\Delta^2}}\left[\sqrt{\frac{\gamma_{\epsilon}}{\gamma}}\,\hat v'_1+\frac{i\omega\Delta}{\gamma^2+\Delta^2}\hat v_2\right]\,.
\ee

Using this modified photodetector output, we compute a full, closed-loop noise model of the system (details in the Appendix). In the ideal feedback limit---i.e., for infinite open-loop gain---the residual force noise
\begin{align}\nonumber
\hat F^{\rm res}|_{{\rm gain}\rightarrow \infty}=\frac{2\hbar G_0\sqrt{\gamma}|\bar a|}{\sqrt{\gamma^2+\Delta^2}}\Bigg\{&\sqrt{\frac{\gamma_{\epsilon}}{\gamma}}
\frac{(\gamma^2-\Delta^2)\hat v_1'+2\gamma\Delta \hat v_2'}{\gamma^2+\Delta^2}\\&-\sqrt{1-\eta}\,\hat n-\frac{i\, \omega\Delta \hat v_2}{\gamma^2+\Delta^2}\Bigg\}\,,
\end{align}
of which the spectral density reads:
\be
S_F^{\rm res} = \frac{4 \hbar^2 G_0^2 \gamma |\bar a|^2}{\gamma^2 + \Delta^2} \left[ \frac{\gamma_\epsilon}{\gamma}+(1-\eta)+\frac{\omega^2 \Delta^2}{(\gamma^2 + \Delta^2)^2} \right].
\ee
By comparison with the thermal force spectrum from a viscous damping model,
$S_{F}^{\rm th} = 4 m \gamma_m k_B T$, we can assign an effective temperature to this
residual force noise as
\be\label{TRes}
T_{\rm eff}^{\rm res} \equiv \frac{S_{F}^{\rm res}}{4 m \gamma_m k_B}.
\ee
In order not to dominate, this residual temperature must be kept below the environmental temperature.

Another interesting result of this closed-loop analysis is that, again for an infinite loop gain, the effective mechanical susceptibility of the resonator becomes
\be
\chi^{-1}_{\rm eff} \rightarrow \chi^{-1}_{\rm eff'} = -m [ \omega^2 + i \gamma_m \omega - (\omega_m^2 + \omega_{\rm os}^2)].
\ee
Comparing this with Eq.\,\eqref{eq_reff}, we see that the damping contribution from the optical spring, $\Gamma_{\rm os}$, is removed. Recall that, for an optical spring with a positive restoring force, we have negative damping: $\Gamma_{\rm os} < 0$ [cf. Eq.\,\eqref{eq_opt}], and it is for this reason that a second optical spring field is usually necessary to make the system stable---the double optical spring scheme\,\cite{Corbitt:2007fk}, discussed in the next section. In our case, if the loop gain is high enough (i.e., if $G_{\rm OL}  \gg |\Gamma_{\rm os}|/|\gamma_m|$), the negative damping will be removed due to the finite response to the mechanical displacement, indicated by Eq.\,\eqref{eq_dIx}, and therefore the system can be stabilized by the positive internal damping of the mechanical system. A practical issue for implementing this is that the required gain could be high in certain applications, and a double optical spring can therefore be used to ease the requirement.

\section{Experimental realization with double optical spring}
In the following, we will detail a proposed experiment, in which a mechanical oscillator with a bare resonance frequency of $\omega_m/2\pi = 100$ Hz is optically stiffened to a new, optomechanical resonance of $\omega_{\rm os}/2\pi \approx 100$ kHz, leading to a commensurate increase in its effective $Q$ factor. This parameter regime is chosen because---due to the low natural loss rate of the resonator---it highlights the long thermal decoherence timescales achievable with such a technique.

Despite the active stabilization effect discussed above, it may be impractical to use a single optical spring due to the very high feedback gains required\footnote{In our example below, using a single optical spring would dictate optical damping $\Gamma_{\rm os}$ on the order of $2\pi \times 1$ kHz. The mechanical damping is $\gamma_m \approx 2\pi \times 10^{-4}$ Hz, and therefore the required gain at the optical spring frequency of 100 kHz is $G_{\rm OL}^{\rm req} \approx 10^{5}$. In practice, obtaining laser amplitude actuation bandwidths above $\sim 1$ MHz is quite challenging, and so it would be difficult to implement a stable loop in this case.}. Instead, we consider a novel approach proposed in Ref.~\cite{Corbitt:2007fk} that uses a second optical spring field to create a passively
stable system. The linear combination of two ${K_{\rm os}}\mbox{s}$, with
one red-detuned and the other blue, can be made to exhibit both positive restoring and damping, resulting in a passively stable spring.
The sum of the two optical spring contributions is thus:
\begin{equation}
K_{\rm os}^{\rm tot} \approx -im\omega \left[ \frac{\gamma_B \omega_{\rm os_B}^2}{(\gamma_B^2 + \Delta_B^2)} - \frac{\gamma_R \omega_{\rm os_R}^2}{(\gamma_R^2 + \Delta_R^2)} \right] + m\omega_{\rm os_B}^2 - m\omega_{\rm os_R}^2
\end{equation}
where $\gamma_B, \gamma_R$ and $\Delta_B, \Delta_R$ are the cavity bandwidth and detuning as seen by the blue and red fields, respectively (note that $\Delta_B < 0$). For a proper choice of these parameters as a function of the ratio $|\omega_{\rm os_B}/\omega_{\rm os_A}| > 1$, the expression in the brackets can be made to vanish, and the effective resonator is stiffened without instability or excess damping\footnote{Note that the expression need not vanish, but only be positive for the resultant resonator to be stable. Furthermore, any positive damping from the optical fields is cold, and therefore does not contribute noise or degrade SNR. We specifically consider the case of zero additional damping, however, since it leads to an effective resonator whose $Q$ is determined solely by the intrinsic damping of the bare mechanical system.}. Additionally, the effect of the feedback discussed above is to suppress the damping contribution from both springs, causing any mismatching of the damping cancellation to be further suppressed. In practice, it may not be trivial to set different bandwidths for two optical fields of macroscopically similar frequency. In this case (i.e., $\gamma_B = \gamma_R \equiv \gamma$), one can still cancel the imaginary terms by choosing the appropriate detunings. In particular, if $|\omega_{\rm os_B}/\omega_{\rm os_A}| = \kappa$, cancellation is obtained for $\Delta_B^2 = (\kappa^2 - 1) \gamma^2 + \kappa^2 \Delta_R^2$.

\begin{table*}[!t]
\caption{A sample set of parameters. These values generate an optical spring with $\omega_{\rm os}/2\pi \approx 100$ kHz and $Q_{\rm eff} \approx 10^9$. The laser powers $P_B$ and $P_R$ refer to the circulating powers, and $Q$ refers to the quality of the bare mechanical system. For the specified geometry, the required finesses are of order $\mathcal{F} \approx 10,000$, compatible with the optical quality of resonators in production today.}\label{tab1}
\begin{center}
\begin{tabular}{|c|cccccccccc|}
\hline
parameter&$m$&$L$&${\omega_m}/{2\pi}$ & $Q$ & ${\gamma_B}/{2\pi}$ &  ${\Delta_B}/{2\pi}$ & $P_B$ & ${\gamma_R}/{2\pi}$ &  ${\Delta_R}/{2\pi}$ & $P_R$\\
\hline
 value & 250 ng & 1 mm & 100 Hz & $10^6$ & 20 MHz & -20 MHz & 390 mW & 4 MHz & 4 MHz & 16 mW\\
\hline
\end{tabular}
\end{center}
\label{t:sample_params}
\end{table*}

A set of sample parameters is given in Table \ref{t:sample_params}. Under these conditions, an oscillator with a resonant frequency of $\omega_{\rm os}/2\pi \approx 100$ kHz and an effective $Q$ of $10^9$ is formed\footnote{This $Q_{\rm eff}$ value is calculated assuming a viscous damping model; the mechanical damping, $\gamma_m$, is fixed, and so, since the optical spring adds no damping, the improvement is given by $Q_{\rm eff} = (\omega_{\rm os} / \omega_m)Q$. Several candidate mechanical resonators are predicted to be better approximated by a structural damping model, in which case the improvement in $Q$ is potentially much greater.}. Such a device can in principle be cooled to its ground state from an environmental temperature of $T \approx 4800$ K (clearly, this should not be attempted, but it serves to illustrate what this technique implies in the context of quantum experiments)! From Eq.\,\eqref{TRes}, we can also calculate the effective temperatures of the residual quantum radiation pressure noise from the two optical spring fields as $T^{\rm res,B}_{\rm eff} = T^{\rm res,R}_{\rm eff} \approx 23$ mK, in the lossless case, or $T^{\rm res,B}_{\rm eff} \approx 84$ K and $T^{\rm res,R}_{\rm eff} \approx 60$ K for realistic losses: 99\% quantum efficiency\,\cite{Eberle:2010fk} and $\epsilon = 30$ ppm\,\cite{Cole:2013kx}. Even in the lossy case, the residual noise temperatures are considerably lower than most target environment temperatures.

\section{Conclusion}
We have proposed a method for creating a tunable effective mechanical resonator with extremely high $Qf$ product. In addition, these resonators can be made to operate in lower frequency bands than current ones of competitive quality, allowing for exceptionally long rethermalization timescales. While the use of optical dilution to mitigate thermal noise has been proposed and demonstrated in the past, we have considered a parameter regime in which the deleterious effects of quantum radiation pressure noise from the strong optical spring fields can be all but eliminated, allowing for greatly unhindered dilution. We feel that the application of this technique holds great promise for any field requiring very-high-Q resonators, including, but not limited to, those of quantum optomechanics and sensitive force measurement.

\section{Acknowledgements}
The authors would like to thank Matt Evans and Sheila Dwyer for several illuminating discussions.

We also gratefully acknowledge support from the National Science Foundation. Specifically: W.~Z.~K. and R.~X.~A. are supported by NSF Grant PHY-0757058; H.~M. and Y.~C. are supported by NSF Grants PHY-1068881 and CAREER Grant PHY-0956189; T.~C. is supported by NSF CAREER Grant PHY-1150531.

\begin{appendix}

\section{Detailed analysis of the system}

In this appendix, we will show some additional details for the derivation of the formulas presented in the main text. We will first consider the ideal case without optical loss and show the leading-order terms in the large bandwidth and detuning limit. Then, we will show the effect of optical loss and next-order correction terms. Finally, we will consider the implementation of feedback and the closed-loop response of the system, which is relevant to actual experimental realization. Our notation here is nearly identical to that in
Ref.\,\cite{Miao:2010zr}.

\subsection{Ideal situation---no optical loss and leading-order terms}\label{sec_ideal}

In this section, we will consider the ideal situation for a typical
optomechanical device,  which has been extensively covered in the literature\,\cite{Marquardt2007, Wilson-Rae2007,
Genes2008a,  Miao:2010zr, Milburn2011}.
We start with the standard Hamiltonian for the canonical optomechanical device,
shown in the dashed box in Fig. 1 of the main text:
\begin{align}\nonumber
\hat{\mathcal{H}}= &\frac{\hat{p}^2}{2m} + \frac{1}{2}m\omega_m^2 \hat{x}^2 + \hbar \omega_c
\hat{a}^{\dag}\hat{a} + \hbar G_0 \hat{x} \hat{a}^\dag \hat{a}\\
& + i \hbar \sqrt{2 \gamma} [\hat{a}_{\rm ext}(t) e^{-i\omega_0 t} \hat{a}^\dag -
\hat{a}_{\rm ext}^\dag(t) e^{i\omega_0 t} \hat{a}]\,.
\end{align}
Here, the first two terms are the free Hamiltonian for the oscillator, with $\omega_m$ being the mechanical
frequency; the third term is the free Hamiltonian for the cavity mode ($\omega_c$ is the cavity resonant
frequency, and $\hat a$ is its annihilation operator satisfying $[\hat a,\,\hat a^{\dag}] = 1)$; the fourth term
describes the interaction between the oscillator and the cavity mode, with $G_0 =\omega_c/L$ being the coupling
strength and $L$ the cavity length; the remaining part is the coupling between the cavity mode with the external
continuum $\hat a_{\rm ext}(t)$, with coupling rate $\gamma$ and $[\hat a_{\rm ext}(t),\,\hat a_{\rm ext}^{\dag}(t')]=\delta(t-t')$, from which one can define the input
operator $\hat a_{\rm in}$ (ingoing before interaction) and output operator $\hat a_{\rm out}$ (outgoing after interaction) through:
\be
\hat a_{\rm in}\equiv \hat a_{\rm ext}(t_-),\quad
\hat a_{\rm out}\equiv \hat a_{\rm ext}(t_+)\,,
\ee
according to the standard input-output formalism\,\cite{Walls2008}.
In the Hamiltonian, we have ignored those terms accounting for the dissipation mechanism of the mechanical oscillator coupling
to its thermal environment. We will later include their effects in the equation of motion for the oscillator.

\subsubsection{Linearized Hamiltonian}
In the experiment, the cavity mode is driven coherently by a laser with a large amplitude at
frequency $\omega_0$. We can therefore study the linearized dynamics by perturbing around the
steady state. In the rotating frame of the laser frequency $\omega_0$,  the corresponding linearized
Hamiltonian for the system reads:
\begin{align}\nonumber
\hat{\mathcal{H}}= &\frac{\hat{p}^2}{2m} + \frac{1}{2}m\omega_m^2 \hat{x}^2 + \hbar \Delta
\hat{a}^{\dag}\hat{a} + \hbar G_0 \hat{x} ( \bar a^*\hat{a}+ \bar a\,\hat{a}^\dag)\\
& + i \hbar \sqrt{2 \gamma} [\hat{a}_{\rm ext}(t) \hat{a}^\dag - \hat{a}_{\rm ext}^\dag(t) \hat{a}]\,.
\end{align}
Here, the cavity detuning is the difference between the cavity resonant frequency and the laser frequency (i.e.,
$\Delta \equiv \omega_c-\omega_0$); $\bar a$ is the steady-state amplitude of the cavity mode, and if we
choose the phase reference such that the steady-state amplitude of the input field, $\bar a_{\rm in}$, is real
and positive, we have
\be
\bar a=\frac{\sqrt{2\gamma}\, \bar a_{\rm in}}{\gamma+i\Delta} = \frac{\sqrt{2 \gamma}}{\gamma + i\Delta} \sqrt{\frac{P_{\rm in}}{\hbar \omega_0}},
 \ee
where $P_{\rm in}$ is the input laser power. These operators in the above
Hamiltonian should be viewed as perturbed parts of the original ones and the quantum state they act on is also transformed
correspondingly. For instance, the input state for $\hat a_{\rm in}$ is
originally a coherent state (for an ideal laser), and now it is the vacuum state $|0\rangle$ with
\be
\langle 0|\hat a_{\rm in}(t)\hat a_{\rm in}^{\dag}(t')|0\rangle = \delta(t-t').
\ee

\subsubsection{Equations of motion}
Given the above Hamiltonian, the cavity mode satisfies the following Heisenberg equation of motion:
\be
\dot{\hat{a}}(t) + (\gamma + i \Delta) \hat{a}(t) = - i G_0\bar a\, \hat{x}(t) + \sqrt{2 \gamma} \,\hat{a}_{\rm in}(t)\,.
\ee
and it is related to the cavity output $\hat a_{\rm out}$ by the standard input-output relation:
\be
\hat a_{\rm out}(t) = -\hat a_{\rm in}(t)+\sqrt{2\gamma}\, \hat a(t).
\ee
Similarly, we can read off the equation of motion for the oscillator:
\be
m [\ddot {\hat x}(t) +\gamma_m \dot {\hat x}(t) +\omega_m^2 \hat x(t)]= \hat F_{\rm rad}(t) + \hat F_{\rm th}(t)\,.
\ee
Here, we have defined the radiation pressure
\be
\hat F_{\rm rad} (t)\equiv - \hbar G_0 [\bar a^*\hat a(t)+\bar a\,\hat a^{\dag}(t)]\,.
 \ee
In addition, we have added the damping term $m\gamma_m \dot {\hat x}(t)$ and the associated thermal fluctuation force $\hat F_{\rm th}$ into
the equation of motion, of which the correlation function is $\langle  \hat F_{\rm th}(t)\hat F_{\rm th}(t')\rangle = 4m \gamma_m k_B T\delta(t-t')$ in the high-temperature limit $k_B T\gg \hbar \omega_m$.

We note here that the equations of motion we have derived are formally identical to those in the classical case. The quantum noise can also be described quasi-classically using a Poisson-statistical approach, however we choose to use the quantum formalism for two reasons: 1) It dramatically simplifies the analysis in cases such as these with multiple loss channels, and 2) despite the absence of explicitly non-classical photonic states, the effects we describe here are fundamentally quantum-mechanical in nature.

\subsubsection{Solution for the cavity mode}
The above linear equations of motion can be solved in the frequency domain. The solution for the
cavity mode reads:
\be
\hat a(\omega) = \frac{G_0 \bar a\,\hat x(\omega) + i\sqrt{2\gamma}\,  \hat a_{\rm in}(\omega)}{\omega-\Delta + i\gamma}\,.
\ee
From this, we can obtain the expression for the radiation pressure:
\be\label{eq_frad}
\hat F_{\rm rad}(\omega) = - K_{\rm os}(\omega)\hat x(\omega) + \hat F_{\rm noise}(\omega)\,.
\ee
We introduce the optical spring coefficient $K_{\rm os}$ as:
\be
K_{\rm os}(\omega)\equiv \frac{2\hbar G_0^2|\bar a|^2 \Delta}{(\omega-\Delta +i\gamma)(\omega+\Delta+i\gamma)}\, ,
\ee
and the quantum radiation pressure noise term as:
\be\label{app_fnoise}
\hat F_{\rm noise}(\omega) \equiv \frac{ 2\hbar G_0|\bar a|\sqrt{\gamma}}{\sqrt{\gamma^2+\Delta^2}}\left[\frac{(\gamma^2+\Delta^2-i\,\gamma\,\omega)\hat v_1+i\,\Delta\,\omega\, \hat v_2}{(\omega-\Delta +i\gamma)(\omega+\Delta+i\gamma)}\right]
\ee
with $\hat v_1\equiv (\hat a_{\rm in}+\hat a^{\dag}_{\rm in})/\sqrt{2}$ and
$\hat v_2\equiv (\hat a_{\rm in}-\hat a_{\rm in}^{\dag})/\sqrt{2}i$ being the vacuum fluctuation of the input amplitude
and phase quadratures, respectively.
The strength of the radiation pressure noise can be quantified by its power spectrum, which is defined through
\be
\langle0 |\hat F^{\dag}_{\rm noise}(\omega)\hat F_{\rm noise}(\omega')|0\rangle_{\rm sym}\equiv \pi\, S_{F}(\omega)\delta(\omega-\omega'),
\ee
where the subscript `sym' denotes for symmetrization and the spectrum is a single-sided one.
Notice that for vacuum input state $\langle 0|\hat v^{\dag}_k(\omega)\hat v_l(\omega')| 0\rangle_{\rm sym}=\pi \,\delta_{kl}\,\delta(\omega-\omega')$, and therefore
\be
S_F(\omega)=\frac{4\hbar^2 G_0^2 |\bar a|^2 \gamma(\gamma^2+\omega^2+\Delta^2)}{[(\omega-\Delta)^2+\gamma^2][(\omega+\Delta)^2+\gamma^2]}\,.
\ee
For the case of large bandwidth and detuning in which we are interested, the above radiation pressure noise can
be approximated as (up to zeroth order of $\omega$):
\be\label{eq_BA_app}
\hat F_{\rm noise} (\omega)\approx-\frac{2\hbar G_0|\bar a|\sqrt{\gamma}}{\sqrt{\gamma^2+\Delta^2}}\hat v_1(\omega)\propto \hat v_1(\omega).
\ee
This indicates that the quantum radiation pressure noise is mostly contributed by fluctuations in the amplitude
quadrature of the input field.
It can be directly measured at the cavity output using a photodetector, as we will see later---this is
why we can evade such noise by feeding back with an appropriate linear filter, which is the
central idea of this work.

\subsubsection{Solution for the mechanical oscillator}
Given the expression for the radiation pressure, we can write down
the solution for the mechanical displacement $\hat x$ as:
\be
\hat x (\omega) =\frac{\hat F_{\rm noise}(\omega)+\hat F_{\rm th}(\omega)}{-m[\omega^2 -\omega_m^2 +i\gamma_m \omega] + K_{\rm os}(\omega)}\,.
\ee
As we can see, the mechanical susceptibility is modified into an effective one due to the optical spring effect. Since
we are focusing on the case of large cavity bandwidth and detuning, the optical spring response $K_{\rm os}$ can be expanded as:
\be
K_{\rm os}\approx -\frac{2\hbar G_0^2|\bar a|^2 \Delta}{\Delta^2+\gamma^2}\left[1 + \frac{2i\gamma \omega}{\Delta^2+\gamma^2}\right]\equiv m \omega_{\rm os}^2 - i\,m \Gamma_{\rm os}\omega\,,
\ee
where $\omega_{\rm os}$ is the optical spring frequency and $\Gamma_{\rm os}$ is the optical damping coefficient.
We can then rewrite the mechanical displacement $\hat x$ as
\be
\hat x (\omega) =  \chi_{\rm eff}(\omega)[\hat F_{\rm noise}(\omega)+\hat F_{\rm th}(\omega)]\,,
\ee
where the effective mechanical susceptibility $\chi_{\rm eff}$ is defined through:
\be\label{eq_reff_app}
\chi^{-1}_{\rm eff}(\omega)\equiv -m[\omega^2+i(\gamma_m +\Gamma_{\rm os})\omega - (\omega_m^2 +\omega_{\rm os}^2)]\,.
\ee
In the negative-detuning case $\Delta<0$, $\omega_{\rm os}$ is positive and real, and the damping $\Gamma_{\rm os}$ is negative; in the positive-detuning case  $\Delta>0$, $\omega_{\rm os}$ is purely imaginary and the damping $\Gamma_{\rm os}$ is positive.
In both cases, the mechanical system is potentially unstable, especially when the intrinsic damping $\gamma_m$ is small as
in our proposed parameter regime. By introducing an additional laser with a different detuning frequency, we can  combine
two optical springs and achieve both positive rigidity and damping---the so-called double optical spring. Such a scheme has
been realized experimentally by Corbitt {\it et al.}\,\cite{Corbitt:2007fk}. We can therefore significantly upshift the mechanical
resonant frequency while keeping the oscillator stable.

\subsubsection{Solution for the cavity output}
From the input-output relation, the cavity output is given by
\be
\hat a_{\rm out}(\omega) =-\frac{\omega-\Delta -i\gamma}{\omega-\Delta+i\gamma}\hat a_{\rm in}(\omega) +\frac{\sqrt{2\gamma}\,G_0\bar a}{\omega-\Delta+i\gamma} \hat x(\omega).
\ee
In the limit of high bandwidth and detuning, we can approximate this as
\be
\hat a_{\rm out}(\omega) = -\frac{\Delta + i\gamma}{\Delta - i\gamma} \hat a_{\rm in}(\omega) - \frac{\sqrt{2 \gamma} G_0 \bar a}{\Delta - i\gamma} \hat x(\omega).
\ee
Similarly, for the classical amplitude at DC, we have the input-output relation:
\be
\bar a_{\rm out} = -\frac{\Delta + i\gamma}{\Delta - i\gamma}\bar a_{\rm in}.
\ee

The photodetector measures the power of the cavity output field:
\begin{align}\nonumber
\hat P_{\rm out}(t)&=|(\bar a_{\rm out}^*+\hat a_{\rm out}^{\dag})(\bar a_{\rm out}+\hat a_{\rm out})|\\&=|\bar a_{\rm out}|^2+\delta \hat P(t)+\hat a_{\rm out}^{\dag}\hat a_{\rm out}.
\end{align}
It contains the classical DC part $|\bar a_{\rm out}|^2$, and the leading-order time-varying
component
\be
\delta \hat P(t)\equiv \bar a_{\rm out}^* \hat a_{\rm out}+\bar a_{\rm out}\hat a_{\rm out}^{\dag}
 \ee
 that we are interested in, which, in the frequency domain, is given by
\be
\delta \hat P(\omega) = \bar a_{\rm out}^* \hat a_{\rm out}(\omega) + \bar a_{\rm out} \hat a_{\rm out}^\dag(\omega) \approx \sqrt{2}\,\bar a_{\rm in}\hat v_1(\omega)\,.
\ee
This means that the photodetector
mostly measures fluctuations in the amplitude quadrature of the input field, which is the main contributor to
the quantum radiation pressure noise felt by the mechanical oscillator as shown by Eq.\,\eqref{eq_BA_app}. Therefore, simply by feeding back the photodetector signal to the mechanical
oscillator, we will be able to evade the quantum radiation pressure noise. The only limitation arises from the optical loss and
the frequency dependence of the radiation pressure noise that we have ignored in Eq.\,\eqref{eq_BA_app} by assuming a large cavity bandwidth and detuning.

\subsection{Realistic situation---optical loss and next-order corrections}\label{sec_realistic}

In this section, we will analyze the effect of optical loss and also the next-order correction---frequency-dependence of the radiation pressure noise as well as non-zero response to the mechanical displacement in the photocurrent---due to finite cavity bandwidth. As mentioned in the main text, the optical loss will decrease the noise cancelation efficiency by introducing vacuum fluctuations---which we denote $\hat a_{\rm in}'$---that are uncorrelated with $\hat a_{\rm in}$. In terms of the equation of motion for the cavity mode, we have
\be
\dot{\hat a}+(\gamma_{\rm tot}+i\Delta)\hat a=-iG_0\bar a \hat x+ \sqrt{2\gamma}\,\hat a_{\rm in}+\sqrt{2\gamma_{\epsilon}}\,\hat a_{\rm in}',
\ee
where $\bar a$ is modified into
\be
\bar a= \frac{\sqrt{2\gamma}\,\bar a_{\rm in}}{\gamma_{\rm tot}+i\Delta}
\ee
and we have introduced
\be
\gamma_{\rm tot}\equiv\gamma+\gamma_{\epsilon}=\gamma+ c\,\epsilon /(4L)
\ee
where $\epsilon$ is the roundtrip power loss factor in the cavity.

\subsubsection{Modification of the radiation pressure}
Correspondingly, this will modify the radiation pressure [cf. Eq.\,\eqref{eq_frad}]:
\be
 \hat F_{\rm rad}(\omega)=-K_{\rm os}(\omega)\hat x(\omega)+\hat F_{\rm noise}(\omega),
 \ee
where
\be
K_{\rm os}=\frac{2\hbar G_0^2|\bar a|^2\Delta}{(\omega-\Delta+i\gamma_{\rm tot})(\omega+\Delta+i\gamma_{\rm tot})}\,,
\ee
and
\begin{align}\nonumber
\hat F_{\rm noise}\equiv&\frac{ 2\hbar G_0  \sqrt{\gamma} |\bar a|}{\sqrt{\gamma_{\rm tot}^2+\Delta^2}}\Bigg\{\frac{(\gamma_{\rm tot}^2+\Delta^2-i\,\gamma_{\rm tot}\,\omega)\hat v_1+i\,\Delta\,\omega\, \hat v_2}{(\omega-\Delta +i\gamma_{\rm tot})(\omega+\Delta+i\gamma_{\rm tot})}
\\ &+\sqrt{\frac{\gamma_{\epsilon}}{\gamma}}\frac{(\gamma_{\rm tot}^2+\Delta^2-i\,\gamma_{\rm tot}\,\omega)\hat v_1'+i\,\Delta\,\omega\, \hat v_2'}{(\omega-\Delta +i\gamma_{\rm tot})(\omega+\Delta+i\gamma_{\rm tot})}\Bigg\}\,,
\end{align}
where $\hat v_1'\equiv (\hat a_{\rm in}'+\hat a_{\rm in}^{'\dag})/\sqrt{2}$ and $\hat v_2'\equiv (\hat a_{\rm in}'-\hat a_{\rm in}^{'\dag})/(\sqrt{2}i)$.

Again for large bandwidth and detuning, and keeping up to the next-order correction---leading order of $\epsilon$ and $\omega$---we obtain:
\be\label{eq_Kos}
K_{\rm os}=-\frac{2\hbar G_0^2|\bar a|^2 \Delta}{\Delta^2+\gamma^2}\left[1-\frac{4\gamma\gamma_{\epsilon}}{\gamma^2+\Delta^2}+\frac{2i\gamma\omega}{\gamma^2+\Delta^2}\right]\,,
\ee
and
\be
\hat F_{\rm noise}=-\frac{2\hbar G_0\sqrt{\gamma}|\bar a|}{\sqrt{\gamma^2+\Delta^2}}\left[\,\hat v_1+\frac{i\omega\Delta}{\gamma^2+\Delta^2}\hat v_2+\sqrt{\frac{\gamma_{\epsilon}}{\gamma}}\,\hat v'_1\right]\,.
\label{eq_fnoisea}
\ee

\subsubsection{Modification of the input-output relation}
Similarly, the input-output relation is also modified into
\begin{align}\nonumber
\hat a_{\rm out}=&-\frac{\omega-\Delta -i(\gamma-\gamma_{\epsilon})}{\omega-\Delta+i\gamma_{\rm  tot}}\hat a_{\rm in}+\frac{2i\sqrt{\gamma\gamma_{\epsilon}}}{\omega-\Delta+i\gamma_{\rm  tot}}\hat a_{\rm in}'\\
&+\frac{\sqrt{2\gamma}\,G_0\bar a}{\omega-\Delta+i\gamma_{\rm tot}} \hat x\,.
\end{align}

\subsubsection{Modification of the photocurrent output}
The exact expression for the AC part of the photocurrent output $\delta \hat P (\omega)$ is
quite complicated, however, in our stated limit, we have
\be
\delta I(\omega)\equiv \sqrt{2}\bar a_{\rm in}\hat v_1(\omega)+ \delta \hat P_{\epsilon}(\omega) +\delta \hat P_{\eta}(\omega)+\delta \hat P_x(\omega)\,,
\ee
where the term $\delta\hat P_{\epsilon}$ contains the vacuum fluctuations $\hat v_{1,2}$ that are associated with optical loss:
\be
\delta \hat P_{\epsilon}(\omega)=\frac{2\sqrt{2\gamma\gamma_{\epsilon}}\bar a_{\rm in}}{\gamma^2+\Delta^2}(\gamma \,\hat v_1'-\Delta\, \hat v_2')\,,
\ee
the additional noise term $\delta \hat P_{\eta}$, due to non-unity quantum inefficiency $\eta$ of the photodetector (keeping to the first order of small $1-\eta$), is
\be
\delta \hat P_{\eta}(\omega)\approx\sqrt{2}\bar a_{\rm in}\sqrt{1-\eta}\,\hat n\,,
\ee
and the term $\delta\hat P_x$ depends on the mechanical displacement:
\be
\delta \hat P_x(\omega)=-\frac{2G_0 |\bar a|^2\Delta(2\gamma_{\epsilon}-i\omega)}{\gamma^2+\Delta^2} \hat x(\omega)\,.
\ee
Therefore, not only is there excess noise from the vacuum fluctuations introduced by the optical loss and non-unity quantum efficiency, but there is also a parasitic sensitivity to mechanical displacement, which is actually associated with the excess radiation pressure [cf. Eq.\,\eqref{eq_fnoisea}], compared with the ideal case [cf. Eq.\,\eqref{eq_BA_app}].

\subsection{Feedback and closed-loop response}
\label{sec_closeloop}

\begin{figure*}[t]
\begin{center}
\includegraphics[width = 180mm]{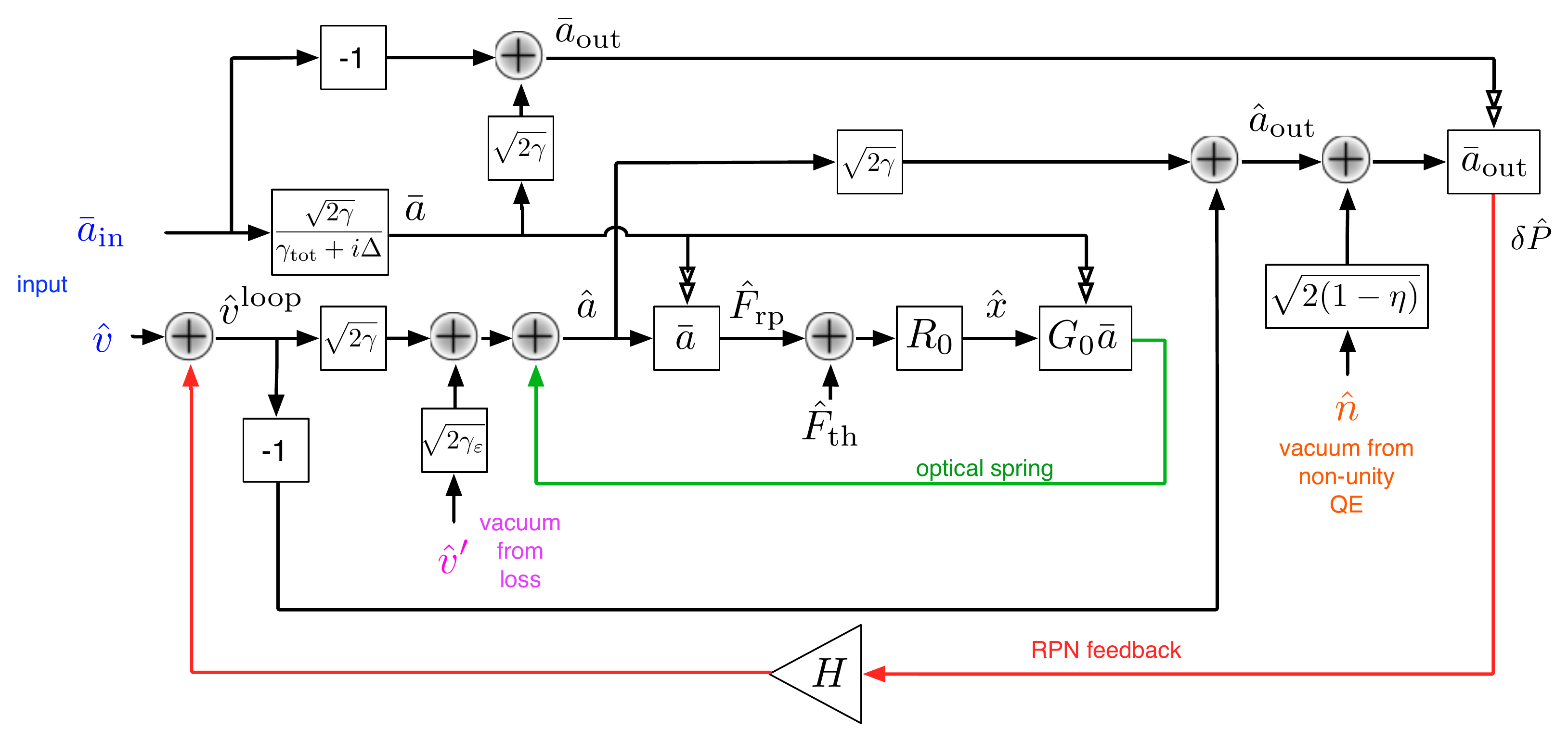}
\caption{A graphical representation of the feedback model described in the text. The effects of the optical spring and the active external feedback are shown explicitly, while the (static) resonance of the classical field $\bar a$ is not. Using the input-output relations, the input fields at left are split into the prompt reflections and the portions that enter the cavity. Then, the leakage fields are summed with the prompt reflections to give the output. The feedback kernel is $H \equiv - K_c(\delta \hat P / \sqrt{2} \bar a_{\rm in})$. Note that the double-headed arrows correspond to the setting of a parameter block, while single-headed arrows denote signal transmission as usual.}
\label{fig:fb_diagram}
\end{center}
\end{figure*}

The radiation pressure noise can be removed either by feedforward (i.e., the photocurrent output is fed forward to the mechanical oscillator as a force), or by feedback (i.e., the photocurrent output is fed back to the input field via an amplitude modulator). Here, we consider the implementation of the feedback scheme. Not only is it more robust against uncertainty in the model transfer functions, but also, as we will show, it can remove the negative damping in the optical spring and stabilize the mechanical oscillator, allowing in principle for a stable {\it single} optical spring.

According to the diagram shown in Fig.\,\ref{fig:fb_diagram}, the
photocurrent output is fed back to an amplitude modulator, which
modulates the amplitude quadrature of the input field. The set of equations for relevant quantities describe such a feedback scheme go as follows, keeping up to the leading order of $\epsilon$ and $\omega$:
\begin{align}
\hat x=&\chi_{\rm eff}(\hat F_{\rm noise}+\hat F_{\rm th})\,,\\
\hat F_{\rm noise}=&-\frac{2\hbar G_0\sqrt{\gamma}|\bar a|}{\sqrt{\gamma^2+\Delta^2}}\left[\hat v_1^{\rm loop}+\frac{i\omega\Delta}{\gamma^2+\Delta^2}\hat v_2+\sqrt{\frac{\gamma_{\epsilon}}{\gamma}}\,\hat v'_1\right]\,,\\\label{dI}
\delta \hat P=&\sqrt{2}\bar a_{\rm in}\hat v_1^{\rm loop}+\delta \hat P_{\epsilon}+\delta \hat P_{\eta}+\delta \hat P_{x}\,,\\\label{v1loop}
\hat v_1^{\rm loop}=&\hat v_1-K_c({\delta \hat P}/\sqrt{2}\bar a_{\rm in})\,.
\end{align}
Here $\hat v_1^{\rm loop}$ is the in-loop amplitude quadrature after the amplitude modulator; $K_c$ is the feedback kernel function and we intentionally leave out the factor $\sqrt{2}\bar a_{\rm in}$ to simplify the equations.

We are interested in the motion of the mechanical oscillator when the feedback is turned on. Solving the above equations leads to
\be
\hat x=\chi_{\rm eff'}(\hat F_{\rm noise'}+\hat F_{\rm th})\,,
\ee
where
\be
\chi^{-1}_{\rm eff'}=\chi^{-1}_{\rm eff}-\frac{4 \hbar G_0^2|\bar a|^2\gamma\Delta(2\gamma_{\epsilon}-i\omega)}{(\gamma^2+\Delta^2)^2}\frac{K_c}{1+K_c}\,,
\ee
and
\begin{align}\nonumber
\hat F_{\rm noise'}=&-\frac{2\hbar G_0\sqrt{\gamma}|\bar a| }{\sqrt{\gamma^2+\Delta^2}}\Bigg[ \frac{1}{1+K_c}\hat v_1
\\\nonumber
&-{\sqrt{\frac{\gamma_{\epsilon}}{\gamma}}}
\left(\frac{K_c-1}{K_c+1}\gamma^2 -\Delta^2\right)\hat v_1' +\frac{2\sqrt{\gamma_{\epsilon}\gamma} \,\Delta K_c}{K_c+1} \hat v_2'\\&-\frac{K_c}{1+K_c}\sqrt{1-\eta}\,\hat n+\frac{i\,\omega\Delta}{\gamma^2+\Delta^2} \hat v_2\Bigg]\,.
\end{align}

\subsubsection{Ideal-feedback limit}
If we make $K_c\rightarrow \infty$, namely, in the ideal feedback limit, we have
\begin{align}\nonumber
\chi^{-1}_{\rm eff'}|_{K_c\rightarrow \infty}&=\chi^{-1}_{\rm eff}-\frac{4 \hbar G_0^2|\bar a|^2\gamma\Delta(2\gamma_{\epsilon}-i\omega)}{(\gamma^2+\Delta^2)^2}\\\nonumber
&=\chi^{-1}_0 +K_{\rm os}-\frac{4 \hbar G_0^2|\bar a|^2\gamma\Delta(2\gamma_{\epsilon}-i\omega)}{(\gamma^2+\Delta^2)^2}\\
&=-m\left[\omega^2+i\gamma_m\omega-(\omega_m^2+\omega_{\rm os}^2)\right]\,,
\end{align}
where we have plugged in the expression for $\chi_{\rm eff}$ [cf. Eq.\,\eqref{eq_reff_app}] and $K_{\rm os}$ [cf. Eq.\,\eqref{eq_Kos}]. Interestingly, the original negative
damping $\Gamma_{\rm os}$ in $K_{\rm os}$ associated with the positive rigidity is canceled out, and the mechanical oscillator
is stabilized. Therefore, using this feedback scheme, the resultant oscillator is stable with a shifted resonant frequency
\be
\omega_m^{\rm new}=\sqrt{\omega_m^2+\omega_{\rm os}^2}\,.
\ee

Now, we quantify the residual radiation pressure noise on
the mechanical oscillator. We have:
\begin{align}\nonumber
\hat F_{\rm noise}'|_{K_c\rightarrow \infty}=\frac{2\hbar G_0\sqrt{\gamma}|\bar a|}{\sqrt{\gamma^2+\Delta^2}}\Bigg\{&\sqrt{\frac{\gamma_{\epsilon}}{\gamma}}\frac{(\gamma^2-\Delta^2)\hat v_1'+2\gamma\Delta \hat v_2'}{\gamma^2+\Delta^2}\\&-\sqrt{1-\eta}\,\hat n-\frac{i\, \omega\Delta \hat v_1}{\gamma^2+\Delta^2} \Bigg\}\,.
\end{align}
The corresponding spectral density reads
\begin{align}
S_{F}^{\rm res}=\frac{4\hbar^2G_0^2\gamma |\bar a|^2}{\gamma^2+\Delta^2}\left[\frac{\gamma_{\epsilon}}{\gamma}+1-\eta+\frac{\omega^2\Delta^2}{(\gamma^2+\Delta^2)^2}\right]\,.
\end{align}
The first term accounts for the effect of the optical loss; the second accounts for non-unity quantum efficiency of the photodetector; the third term accounts for a finite cavity bandwidth.

\subsection{Proposed experimental setup}

\begin{figure}[h]
\begin{center}
\includegraphics[width=\columnwidth]{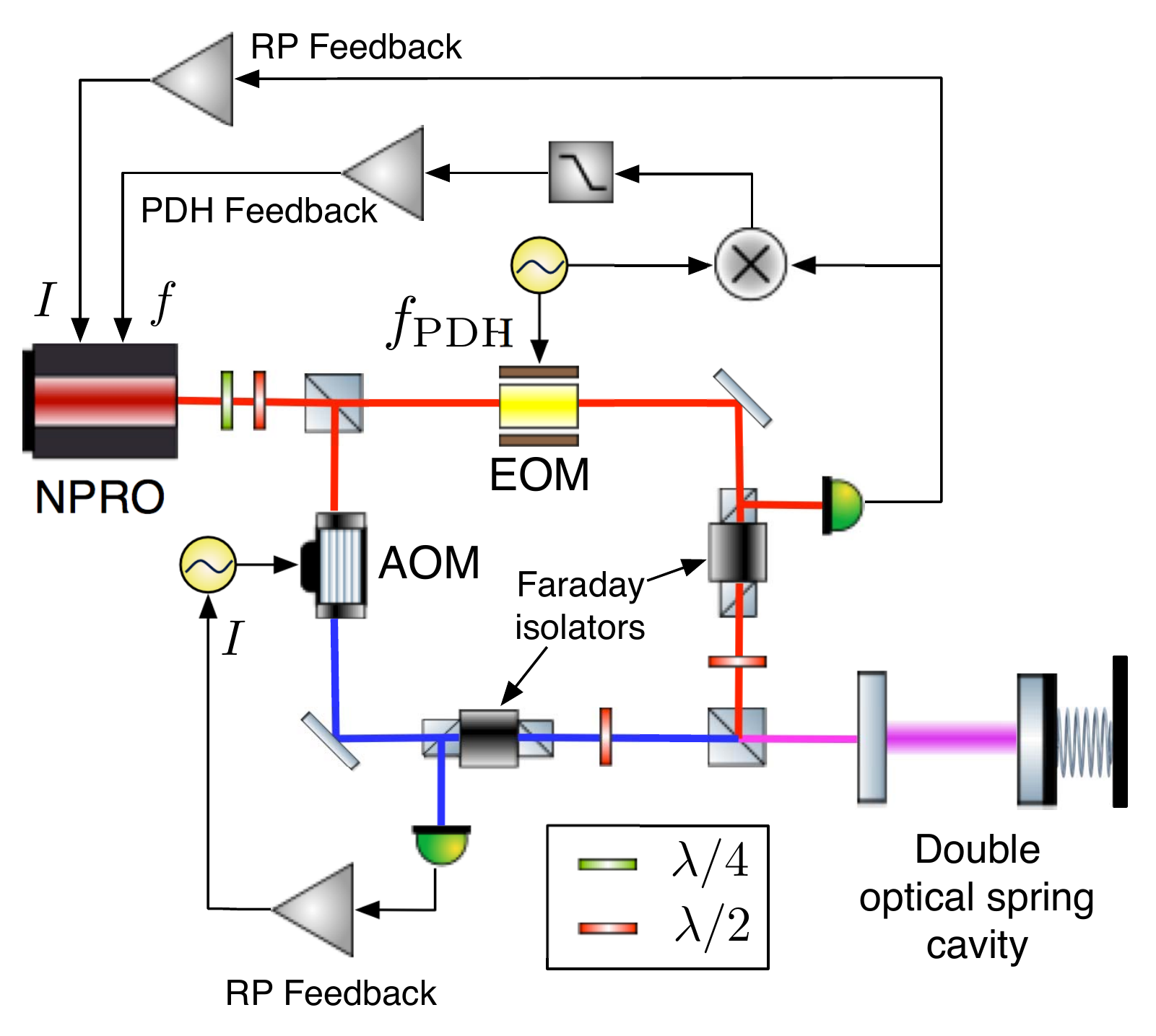}
\caption{The proposed experimental setup. A laser beam is split and one path is upshifted in frequency, allowing for independent control of the detuning of each field. Each beam's intensity can also be controlled by feeding back to the laser or the modulator, and these channels are used for the radiation pressure noise feedback. A Pound-Drever-Hall locking scheme is used to set the operating point before strengthening the stable optical spring.}
\label{fig:exp_setup}
\end{center}
\end{figure}

While the technique described in this paper is quite general, a possible experimental layout is shown in Fig.\,\ref{fig:exp_setup}. A laser's frequency is stabilized to the optical spring cavity length using the Pound-Drever-Hall (PDH)\cite{Drever:1983fk} locking technique. The laser is then detuned from the resonance by injecting an offset into the error point of the control loop. A second beam is picked off from the main laser and upshifted in frequency by an acousto-optic modulator (AOM). Once the detuning is set properly for both beams, the input power is ramped until the optical spring reaches the desired strength. At this point, the PDH frequency feedback to the laser can be disengaged, and---provided the mechanical and laser frequency stability are sufficiently high---the resonator is trapped in a passively stable potential by the optical spring forces.

Finally, the radiation pressure noise feedback described above is engaged, with the primary beam's signal fed back to the laser amplitude, and the secondary's to the AOM drive amplitude. The EOM, which imparts the phase modulation sidebands necessary for PDH lock, is also disengaged so as not to couple extra uncorrelated vacuum noise into the readout. In this operational configuration, the quantum radiation pressure noise is very strongly suppressed, limited only by the parasitic loss and finite-bandwidth effects detailed above. Classical laser amplitude noise---which is indistinguishable from its quantum counterpart here---is also suppressed by the loop.

The only remaining potential issues are laser frequency stability and drift of the mechanical system, which can drive the optical spring fields away from their optimal detunings. The former can be avoided using pre-stabilization (e.g., by locking the laser to an external frequency reference). The latter is not as simple to avoid, and will depend on the mechanical system in question.  If necessary, a very weak PDH lock can be maintained using a low-frequency servo to ensure DC stability of the operating point. In this case, it may be possible to use weak enough control sideband fields that the RPN readout is still limited by the finite losses and bandwidth.

\end{appendix}

\bibliography{noise-free_OS}

\begin{thebibliography}{41}
\expandafter\ifx\csname natexlab\endcsname\relax\def\natexlab#1{#1}\fi
\expandafter\ifx\csname bibnamefont\endcsname\relax
  \def\bibnamefont#1{#1}\fi
\expandafter\ifx\csname bibfnamefont\endcsname\relax
  \def\bibfnamefont#1{#1}\fi
\expandafter\ifx\csname citenamefont\endcsname\relax
  \def\citenamefont#1{#1}\fi
\expandafter\ifx\csname url\endcsname\relax
  \def\url#1{\texttt{#1}}\fi
\expandafter\ifx\csname urlprefix\endcsname\relax\def\urlprefix{URL }\fi
\providecommand{\bibinfo}[2]{#2}
\providecommand{\eprint}[2][]{\url{#2}}

\bibitem[{\citenamefont{Kippenberg and Vahala}(2008)}]{Kippenberg2008}
\bibinfo{author}{\bibfnamefont{T.~J.} \bibnamefont{Kippenberg}}
  \bibnamefont{and} \bibinfo{author}{\bibfnamefont{K.~J.}
  \bibnamefont{Vahala}}, \bibinfo{journal}{Science}
  \textbf{\bibinfo{volume}{321}}, \bibinfo{pages}{1172} (\bibinfo{year}{2008}),
  \urlprefix\url{http://www.sciencemag.org/cgi/content/abstract/321/5893/1172}.

\bibitem[{\citenamefont{Marquardt and Girvin}(2009)}]{Marquardt:2009fk}
\bibinfo{author}{\bibfnamefont{F.}~\bibnamefont{Marquardt}} \bibnamefont{and}
  \bibinfo{author}{\bibfnamefont{S.~M.} \bibnamefont{Girvin}},
  \bibinfo{journal}{Physics} \textbf{\bibinfo{volume}{2}}, \bibinfo{pages}{40}
  (\bibinfo{year}{2009}),
  \urlprefix\url{http://link.aps.org/doi/10.1103/Physics.2.40}.

\bibitem[{\citenamefont{Aspelmeyer et~al.}(2012)\citenamefont{Aspelmeyer,
  Meystre, and Schwab}}]{Aspelmeyer_review}
\bibinfo{author}{\bibfnamefont{M.}~\bibnamefont{Aspelmeyer}},
  \bibinfo{author}{\bibfnamefont{P.}~\bibnamefont{Meystre}}, \bibnamefont{and}
  \bibinfo{author}{\bibfnamefont{K.}~\bibnamefont{Schwab}},
  \bibinfo{journal}{Physics Today} \textbf{\bibinfo{volume}{65}},
  \bibinfo{pages}{29} (\bibinfo{year}{2012}).

\bibitem[{\citenamefont{Braginsky and Khalili}(1992)}]{BK92}
\bibinfo{author}{\bibfnamefont{V.~B.} \bibnamefont{Braginsky}}
  \bibnamefont{and} \bibinfo{author}{\bibfnamefont{F.~Y.}
  \bibnamefont{Khalili}}, \emph{\bibinfo{title}{Quantum Measurement}}
  (\bibinfo{publisher}{Cambridge University Press}, \bibinfo{year}{1992}).

\bibitem[{\citenamefont{Abbott et~al.}(2009)\citenamefont{Abbott, Abbott,
  Adhikari, Ajith, Allen, Allen, Amin, Anderson, Anderson, Arain
  et~al.}}]{LIGO_review}
\bibinfo{author}{\bibfnamefont{B.~P.} \bibnamefont{Abbott}},
  \bibinfo{author}{\bibfnamefont{R.}~\bibnamefont{Abbott}},
  \bibinfo{author}{\bibfnamefont{R.}~\bibnamefont{Adhikari}},
  \bibinfo{author}{\bibfnamefont{P.}~\bibnamefont{Ajith}},
  \bibinfo{author}{\bibfnamefont{B.}~\bibnamefont{Allen}},
  \bibinfo{author}{\bibfnamefont{G.}~\bibnamefont{Allen}},
  \bibinfo{author}{\bibfnamefont{R.~S.} \bibnamefont{Amin}},
  \bibinfo{author}{\bibfnamefont{S.~B.} \bibnamefont{Anderson}},
  \bibinfo{author}{\bibfnamefont{W.~G.} \bibnamefont{Anderson}},
  \bibinfo{author}{\bibfnamefont{M.~A.} \bibnamefont{Arain}},
  \bibnamefont{et~al.}, \bibinfo{journal}{Reports on Progress in Physics}
  \textbf{\bibinfo{volume}{72}}, \bibinfo{pages}{076901}
  (\bibinfo{year}{2009}),
  \urlprefix\url{http://stacks.iop.org/0034-4885/72/i=7/a=076901}.

\bibitem[{\citenamefont{Teufel et~al.}(2011)\citenamefont{Teufel, Donner, Li,
  Harlow, Allman, Cicak, Sirois, Whittaker, Lehnert, and
  Simmonds}}]{Teufel2011}
\bibinfo{author}{\bibfnamefont{J.~D.} \bibnamefont{Teufel}},
  \bibinfo{author}{\bibfnamefont{T.}~\bibnamefont{Donner}},
  \bibinfo{author}{\bibfnamefont{D.}~\bibnamefont{Li}},
  \bibinfo{author}{\bibfnamefont{J.~W.} \bibnamefont{Harlow}},
  \bibinfo{author}{\bibfnamefont{M.~S.} \bibnamefont{Allman}},
  \bibinfo{author}{\bibfnamefont{K.}~\bibnamefont{Cicak}},
  \bibinfo{author}{\bibfnamefont{A.~J.} \bibnamefont{Sirois}},
  \bibinfo{author}{\bibfnamefont{J.~D.} \bibnamefont{Whittaker}},
  \bibinfo{author}{\bibfnamefont{K.~W.} \bibnamefont{Lehnert}},
  \bibnamefont{and} \bibinfo{author}{\bibfnamefont{R.~W.}
  \bibnamefont{Simmonds}}, \bibinfo{journal}{Nature}
  \textbf{\bibinfo{volume}{475}}, \bibinfo{pages}{359} (\bibinfo{year}{2011}),
  ISSN \bibinfo{issn}{0028-0836},
  \urlprefix\url{http://dx.doi.org/10.1038/nature10261}.

\bibitem[{\citenamefont{O'Connell}(2010)}]{OConnell2010}
\bibinfo{author}{\bibfnamefont{A.~D.} \bibnamefont{O'Connell}},
  \bibinfo{journal}{Nature} \textbf{\bibinfo{volume}{464}},
  \bibinfo{pages}{697} (\bibinfo{year}{2010}),
  \urlprefix\url{http://dx.doi.org/10.1038/nature08967}.

\bibitem[{\citenamefont{Chan et~al.}(2011)\citenamefont{Chan, Alegre,
  Safavi-Naeini, Hill, Krause, Groblacher, Aspelmeyer, and
  Painter}}]{Chan:2011uq}
\bibinfo{author}{\bibfnamefont{J.}~\bibnamefont{Chan}},
  \bibinfo{author}{\bibfnamefont{T.~P.~M.} \bibnamefont{Alegre}},
  \bibinfo{author}{\bibfnamefont{A.~H.} \bibnamefont{Safavi-Naeini}},
  \bibinfo{author}{\bibfnamefont{J.~T.} \bibnamefont{Hill}},
  \bibinfo{author}{\bibfnamefont{A.}~\bibnamefont{Krause}},
  \bibinfo{author}{\bibfnamefont{S.}~\bibnamefont{Groblacher}},
  \bibinfo{author}{\bibfnamefont{M.}~\bibnamefont{Aspelmeyer}},
  \bibnamefont{and} \bibinfo{author}{\bibfnamefont{O.}~\bibnamefont{Painter}},
  \bibinfo{journal}{Nature} \textbf{\bibinfo{volume}{478}}, \bibinfo{pages}{89}
  (\bibinfo{year}{2011}).

\bibitem[{\citenamefont{Weis et~al.}(2010)\citenamefont{Weis, Rivi{\`e}re,
  Del{\'e}glise, Gavartin, Arcizet, Schliesser, and Kippenberg}}]{Weis10122010}
\bibinfo{author}{\bibfnamefont{S.}~\bibnamefont{Weis}},
  \bibinfo{author}{\bibfnamefont{R.}~\bibnamefont{Rivi{\`e}re}},
  \bibinfo{author}{\bibfnamefont{S.}~\bibnamefont{Del{\'e}glise}},
  \bibinfo{author}{\bibfnamefont{E.}~\bibnamefont{Gavartin}},
  \bibinfo{author}{\bibfnamefont{O.}~\bibnamefont{Arcizet}},
  \bibinfo{author}{\bibfnamefont{A.}~\bibnamefont{Schliesser}},
  \bibnamefont{and} \bibinfo{author}{\bibfnamefont{T.~J.}
  \bibnamefont{Kippenberg}}, \bibinfo{journal}{Science}
  \textbf{\bibinfo{volume}{330}}, \bibinfo{pages}{1520} (\bibinfo{year}{2010}),
  \eprint{http://www.sciencemag.org/content/330/6010/1520.full.pdf},
  \urlprefix\url{http://www.sciencemag.org/content/330/6010/1520.abstract}.

\bibitem[{\citenamefont{Safavi-Naeini et~al.}(2011)\citenamefont{Safavi-Naeini,
  Alegre, Chan, Eichenfield, Winger, Lin, Hill, Chang, and
  Painter}}]{Safavi-Naeini:2011kx}
\bibinfo{author}{\bibfnamefont{A.~H.} \bibnamefont{Safavi-Naeini}},
  \bibinfo{author}{\bibfnamefont{T.~P.~M.} \bibnamefont{Alegre}},
  \bibinfo{author}{\bibfnamefont{J.}~\bibnamefont{Chan}},
  \bibinfo{author}{\bibfnamefont{M.}~\bibnamefont{Eichenfield}},
  \bibinfo{author}{\bibfnamefont{M.}~\bibnamefont{Winger}},
  \bibinfo{author}{\bibfnamefont{Q.}~\bibnamefont{Lin}},
  \bibinfo{author}{\bibfnamefont{J.~T.} \bibnamefont{Hill}},
  \bibinfo{author}{\bibfnamefont{D.~E.} \bibnamefont{Chang}}, \bibnamefont{and}
  \bibinfo{author}{\bibfnamefont{O.}~\bibnamefont{Painter}},
  \bibinfo{journal}{Nature} \textbf{\bibinfo{volume}{472}}, \bibinfo{pages}{69}
  (\bibinfo{year}{2011}).

\bibitem[{\citenamefont{Boller et~al.}(1991)\citenamefont{Boller, Imamolu, and
  Harris}}]{Boller:1991yq}
\bibinfo{author}{\bibfnamefont{K.-J.} \bibnamefont{Boller}},
  \bibinfo{author}{\bibfnamefont{A.}~\bibnamefont{Imamolu}}, \bibnamefont{and}
  \bibinfo{author}{\bibfnamefont{S.~E.} \bibnamefont{Harris}},
  \bibinfo{journal}{Phys. Rev. Lett.} \textbf{\bibinfo{volume}{66}},
  \bibinfo{pages}{2593} (\bibinfo{year}{1991}),
  \urlprefix\url{http://link.aps.org/doi/10.1103/PhysRevLett.66.2593}.

\bibitem[{\citenamefont{Harris}(1997)}]{Harris:1997ly}
\bibinfo{author}{\bibfnamefont{S.~E.} \bibnamefont{Harris}},
  \bibinfo{journal}{Physics Today} \textbf{\bibinfo{volume}{50}},
  \bibinfo{pages}{36} (\bibinfo{year}{1997}),
  \urlprefix\url{http://link.aip.org/link/?PTO/50/36/1}.

\bibitem[{\citenamefont{Kimble et~al.}(2001)\citenamefont{Kimble, Levin,
  Matsko, Thorne, and Vyatchanin}}]{Kimble:2001kx}
\bibinfo{author}{\bibfnamefont{H.~J.} \bibnamefont{Kimble}},
  \bibinfo{author}{\bibfnamefont{Y.}~\bibnamefont{Levin}},
  \bibinfo{author}{\bibfnamefont{A.~B.} \bibnamefont{Matsko}},
  \bibinfo{author}{\bibfnamefont{K.~S.} \bibnamefont{Thorne}},
  \bibnamefont{and} \bibinfo{author}{\bibfnamefont{S.~P.}
  \bibnamefont{Vyatchanin}}, \bibinfo{journal}{Phys. Rev. D}
  \textbf{\bibinfo{volume}{65}}, \bibinfo{pages}{022002}
  (\bibinfo{year}{2001}),
  \urlprefix\url{http://link.aps.org/doi/10.1103/PhysRevD.65.022002}.

\bibitem[{\citenamefont{Ma et~al.}(2013)\citenamefont{Ma, Danilishin, Korth,
  Miao, Chen, and Zhao}}]{Yiqiu2012}
\bibinfo{author}{\bibfnamefont{Y.}~\bibnamefont{Ma}},
  \bibinfo{author}{\bibfnamefont{S.}~\bibnamefont{Danilishin}},
  \bibinfo{author}{\bibfnamefont{W.~Z.} \bibnamefont{Korth}},
  \bibinfo{author}{\bibfnamefont{H.}~\bibnamefont{Miao}},
  \bibinfo{author}{\bibfnamefont{Y.}~\bibnamefont{Chen}}, \bibnamefont{and}
  \bibinfo{author}{\bibfnamefont{C.}~\bibnamefont{Zhao}}, \bibinfo{journal}{In
  preparation}  (\bibinfo{year}{2013}).

\bibitem[{\citenamefont{Saulson}(1990)}]{Saulson:1990qy}
\bibinfo{author}{\bibfnamefont{P.~R.} \bibnamefont{Saulson}},
  \bibinfo{journal}{Phys. Rev. D} \textbf{\bibinfo{volume}{42}},
  \bibinfo{pages}{2437} (\bibinfo{year}{1990}),
  \urlprefix\url{http://link.aps.org/doi/10.1103/PhysRevD.42.2437}.

\bibitem[{\citenamefont{McGuigan et~al.}(1978)\citenamefont{McGuigan, Lam,
  Gram, Hoffman, Douglass, and Gutche}}]{McGuigan:1978fj}
\bibinfo{author}{\bibfnamefont{D.~F.} \bibnamefont{McGuigan}},
  \bibinfo{author}{\bibfnamefont{C.~C.} \bibnamefont{Lam}},
  \bibinfo{author}{\bibfnamefont{R.~Q.} \bibnamefont{Gram}},
  \bibinfo{author}{\bibfnamefont{A.~W.} \bibnamefont{Hoffman}},
  \bibinfo{author}{\bibfnamefont{D.~H.} \bibnamefont{Douglass}},
  \bibnamefont{and} \bibinfo{author}{\bibfnamefont{H.~W.}
  \bibnamefont{Gutche}}, \bibinfo{journal}{Journal of Low Temperature Physics}
  \textbf{\bibinfo{volume}{30}}, \bibinfo{pages}{621} (\bibinfo{year}{1978}),
  ISSN \bibinfo{issn}{0022-2291}, \bibinfo{note}{10.1007/BF00116202},
  \urlprefix\url{http://dx.doi.org/10.1007/BF00116202}.

\bibitem[{\citenamefont{Ageev et~al.}(2004)\citenamefont{Ageev, Palmer, Felice,
  Penn, and Saulson}}]{Ageev:2004yq}
\bibinfo{author}{\bibfnamefont{A.}~\bibnamefont{Ageev}},
  \bibinfo{author}{\bibfnamefont{B.~C.} \bibnamefont{Palmer}},
  \bibinfo{author}{\bibfnamefont{A.~D.} \bibnamefont{Felice}},
  \bibinfo{author}{\bibfnamefont{S.~D.} \bibnamefont{Penn}}, \bibnamefont{and}
  \bibinfo{author}{\bibfnamefont{P.~R.} \bibnamefont{Saulson}},
  \bibinfo{journal}{Classical and Quantum Gravity}
  \textbf{\bibinfo{volume}{21}}, \bibinfo{pages}{3887} (\bibinfo{year}{2004}),
  \urlprefix\url{http://stacks.iop.org/0264-9381/21/i=16/a=004}.

\bibitem[{\citenamefont{Mihailovich and MacDonald}(1995)}]{Mihailovich:1995fk}
\bibinfo{author}{\bibfnamefont{R.}~\bibnamefont{Mihailovich}} \bibnamefont{and}
  \bibinfo{author}{\bibfnamefont{N.}~\bibnamefont{MacDonald}},
  \bibinfo{journal}{Sensors and Actuators A: Physical}
  \textbf{\bibinfo{volume}{50}}, \bibinfo{pages}{199 } (\bibinfo{year}{1995}),
  ISSN \bibinfo{issn}{0924-4247},
  \urlprefix\url{http://www.sciencedirect.com/science/article/pii/0924424795010807}.

\bibitem[{\citenamefont{Cole et~al.}(2011)\citenamefont{Cole, Wilson-Rae,
  Werbach, Vanner, and Aspelmeyer}}]{Cole:2011rt}
\bibinfo{author}{\bibfnamefont{G.~D.} \bibnamefont{Cole}},
  \bibinfo{author}{\bibfnamefont{I.}~\bibnamefont{Wilson-Rae}},
  \bibinfo{author}{\bibfnamefont{K.}~\bibnamefont{Werbach}},
  \bibinfo{author}{\bibfnamefont{M.~R.} \bibnamefont{Vanner}},
  \bibnamefont{and}
  \bibinfo{author}{\bibfnamefont{M.}~\bibnamefont{Aspelmeyer}},
  \bibinfo{journal}{Nat. Commun.} \textbf{\bibinfo{volume}{2}},
  \bibinfo{pages}{231} (\bibinfo{year}{2011}).

\bibitem[{\citenamefont{Zener}(1938)}]{Zener:1938zr}
\bibinfo{author}{\bibfnamefont{C.}~\bibnamefont{Zener}},
  \bibinfo{journal}{Phys. Rev.} \textbf{\bibinfo{volume}{53}},
  \bibinfo{pages}{90} (\bibinfo{year}{1938}),
  \urlprefix\url{http://link.aps.org/doi/10.1103/PhysRev.53.90}.

\bibitem[{\citenamefont{Akhiezer}(1939)}]{Akhiezer:1939ly}
\bibinfo{author}{\bibfnamefont{A.}~\bibnamefont{Akhiezer}},
  \bibinfo{journal}{J. Phys. (Moscow)} \textbf{\bibinfo{volume}{1}},
  \bibinfo{pages}{277} (\bibinfo{year}{1939}).

\bibitem[{\citenamefont{Harry et~al.}(2002)\citenamefont{Harry, Gretarsson,
  Saulson, Kittelberger, Penn, Startin, Rowan, Fejer, Crooks, Cagnoli
  et~al.}}]{Harry:2002ve}
\bibinfo{author}{\bibfnamefont{G.~M.} \bibnamefont{Harry}},
  \bibinfo{author}{\bibfnamefont{A.~M.} \bibnamefont{Gretarsson}},
  \bibinfo{author}{\bibfnamefont{P.~R.} \bibnamefont{Saulson}},
  \bibinfo{author}{\bibfnamefont{S.~E.} \bibnamefont{Kittelberger}},
  \bibinfo{author}{\bibfnamefont{S.~D.} \bibnamefont{Penn}},
  \bibinfo{author}{\bibfnamefont{W.~J.} \bibnamefont{Startin}},
  \bibinfo{author}{\bibfnamefont{S.}~\bibnamefont{Rowan}},
  \bibinfo{author}{\bibfnamefont{M.~M.} \bibnamefont{Fejer}},
  \bibinfo{author}{\bibfnamefont{D.~R.~M.} \bibnamefont{Crooks}},
  \bibinfo{author}{\bibfnamefont{G.}~\bibnamefont{Cagnoli}},
  \bibnamefont{et~al.}, \bibinfo{journal}{Classical and Quantum Gravity}
  \textbf{\bibinfo{volume}{19}}, \bibinfo{pages}{897} (\bibinfo{year}{2002}),
  \urlprefix\url{http://stacks.iop.org/0264-9381/19/i=5/a=305}.

\bibitem[{\citenamefont{Braginsky et~al.}(1997)\citenamefont{Braginsky,
  Gorodetsky, and Khalili}}]{Braginsky1997}
\bibinfo{author}{\bibfnamefont{V.}~\bibnamefont{Braginsky}},
  \bibinfo{author}{\bibfnamefont{M.}~\bibnamefont{Gorodetsky}},
  \bibnamefont{and} \bibinfo{author}{\bibfnamefont{F.}~\bibnamefont{Khalili}},
  \bibinfo{journal}{Physics Letters A} \textbf{\bibinfo{volume}{232}},
  \bibinfo{pages}{340} (\bibinfo{year}{1997}), ISSN \bibinfo{issn}{0375-9601},
  \urlprefix\url{http://www.sciencedirect.com/science/article/pii/S0375960197004131}.

\bibitem[{\citenamefont{Braginsky and Khalili}(1999)}]{Braginsky:1999ys}
\bibinfo{author}{\bibfnamefont{V.}~\bibnamefont{Braginsky}} \bibnamefont{and}
  \bibinfo{author}{\bibfnamefont{F.}~\bibnamefont{Khalili}},
  \bibinfo{journal}{Physics Letters A} \textbf{\bibinfo{volume}{257}},
  \bibinfo{pages}{241 } (\bibinfo{year}{1999}), ISSN \bibinfo{issn}{0375-9601},
  \urlprefix\url{http://www.sciencedirect.com/science/article/pii/S0375960199003370}.

\bibitem[{\citenamefont{Buonanno and Chen}(2002)}]{Buonanno2002}
\bibinfo{author}{\bibfnamefont{A.}~\bibnamefont{Buonanno}} \bibnamefont{and}
  \bibinfo{author}{\bibfnamefont{Y.}~\bibnamefont{Chen}},
  \bibinfo{journal}{Phys. Rev. D} \textbf{\bibinfo{volume}{65}},
  \bibinfo{pages}{042001} (\bibinfo{year}{2002}),
  \urlprefix\url{http://link.aps.org/doi/10.1103/PhysRevD.65.042001}.

\bibitem[{\citenamefont{Corbitt
  et~al.}(2007{\natexlab{a}})\citenamefont{Corbitt, Chen, Innerhofer,
  M{\"u}ller-Ebhardt, Ottaway, Rehbein, Sigg, Whitcomb, Wipf, and
  Mavalvala}}]{Corbitt:2007fk}
\bibinfo{author}{\bibfnamefont{T.}~\bibnamefont{Corbitt}},
  \bibinfo{author}{\bibfnamefont{Y.}~\bibnamefont{Chen}},
  \bibinfo{author}{\bibfnamefont{E.}~\bibnamefont{Innerhofer}},
  \bibinfo{author}{\bibfnamefont{H.}~\bibnamefont{M{\"u}ller-Ebhardt}},
  \bibinfo{author}{\bibfnamefont{D.}~\bibnamefont{Ottaway}},
  \bibinfo{author}{\bibfnamefont{H.}~\bibnamefont{Rehbein}},
  \bibinfo{author}{\bibfnamefont{D.}~\bibnamefont{Sigg}},
  \bibinfo{author}{\bibfnamefont{S.}~\bibnamefont{Whitcomb}},
  \bibinfo{author}{\bibfnamefont{C.}~\bibnamefont{Wipf}}, \bibnamefont{and}
  \bibinfo{author}{\bibfnamefont{N.}~\bibnamefont{Mavalvala}},
  \bibinfo{journal}{Phys. Rev. Lett.} \textbf{\bibinfo{volume}{98}},
  \bibinfo{pages}{150802} (\bibinfo{year}{2007}{\natexlab{a}}),
  \urlprefix\url{http://link.aps.org/doi/10.1103/PhysRevLett.98.150802}.

\bibitem[{\citenamefont{Corbitt
  et~al.}(2007{\natexlab{b}})\citenamefont{Corbitt, Wipf, Bodiya, Ottaway,
  Sigg, Smith, Whitcomb, and Mavalvala}}]{Corbitt:2007lr}
\bibinfo{author}{\bibfnamefont{T.}~\bibnamefont{Corbitt}},
  \bibinfo{author}{\bibfnamefont{C.}~\bibnamefont{Wipf}},
  \bibinfo{author}{\bibfnamefont{T.}~\bibnamefont{Bodiya}},
  \bibinfo{author}{\bibfnamefont{D.}~\bibnamefont{Ottaway}},
  \bibinfo{author}{\bibfnamefont{D.}~\bibnamefont{Sigg}},
  \bibinfo{author}{\bibfnamefont{N.}~\bibnamefont{Smith}},
  \bibinfo{author}{\bibfnamefont{S.}~\bibnamefont{Whitcomb}}, \bibnamefont{and}
  \bibinfo{author}{\bibfnamefont{N.}~\bibnamefont{Mavalvala}},
  \bibinfo{journal}{Phys. Rev. Lett.} \textbf{\bibinfo{volume}{99}},
  \bibinfo{pages}{160801} (\bibinfo{year}{2007}{\natexlab{b}}),
  \urlprefix\url{http://link.aps.org/doi/10.1103/PhysRevLett.99.160801}.

\bibitem[{\citenamefont{Di~Virgilio et~al.}(2006)\citenamefont{Di~Virgilio,
  Barsotti, Braccini, Bradaschia, Cella, Corda, Dattilo, Ferrante, Fidecaro,
  Fiori et~al.}}]{Di-Virgilio:2006fk}
\bibinfo{author}{\bibfnamefont{A.}~\bibnamefont{Di~Virgilio}},
  \bibinfo{author}{\bibfnamefont{L.}~\bibnamefont{Barsotti}},
  \bibinfo{author}{\bibfnamefont{S.}~\bibnamefont{Braccini}},
  \bibinfo{author}{\bibfnamefont{C.}~\bibnamefont{Bradaschia}},
  \bibinfo{author}{\bibfnamefont{G.}~\bibnamefont{Cella}},
  \bibinfo{author}{\bibfnamefont{C.}~\bibnamefont{Corda}},
  \bibinfo{author}{\bibfnamefont{V.}~\bibnamefont{Dattilo}},
  \bibinfo{author}{\bibfnamefont{I.}~\bibnamefont{Ferrante}},
  \bibinfo{author}{\bibfnamefont{F.}~\bibnamefont{Fidecaro}},
  \bibinfo{author}{\bibfnamefont{I.}~\bibnamefont{Fiori}},
  \bibnamefont{et~al.}, \bibinfo{journal}{Phys. Rev. A}
  \textbf{\bibinfo{volume}{74}}, \bibinfo{pages}{013813}
  (\bibinfo{year}{2006}),
  \urlprefix\url{http://link.aps.org/doi/10.1103/PhysRevA.74.013813}.

\bibitem[{\citenamefont{Chang et~al.}(2012)\citenamefont{Chang, Ni, Painter,
  and Kimble}}]{Chang:2012fj}
\bibinfo{author}{\bibfnamefont{D.~E.} \bibnamefont{Chang}},
  \bibinfo{author}{\bibfnamefont{K.-K.} \bibnamefont{Ni}},
  \bibinfo{author}{\bibfnamefont{O.}~\bibnamefont{Painter}}, \bibnamefont{and}
  \bibinfo{author}{\bibfnamefont{H.~J.} \bibnamefont{Kimble}},
  \bibinfo{journal}{New Journal of Physics} \textbf{\bibinfo{volume}{14}},
  \bibinfo{pages}{045002} (\bibinfo{year}{2012}),
  \urlprefix\url{http://stacks.iop.org/1367-2630/14/i=4/a=045002}.

\bibitem[{\citenamefont{Ni et~al.}(2012)\citenamefont{Ni, Norte, Wilson, Hood,
  Chang, Painter, and Kimble}}]{Ni:2012qy}
\bibinfo{author}{\bibfnamefont{K.-K.} \bibnamefont{Ni}},
  \bibinfo{author}{\bibfnamefont{R.}~\bibnamefont{Norte}},
  \bibinfo{author}{\bibfnamefont{D.~J.} \bibnamefont{Wilson}},
  \bibinfo{author}{\bibfnamefont{J.~D.} \bibnamefont{Hood}},
  \bibinfo{author}{\bibfnamefont{D.~E.} \bibnamefont{Chang}},
  \bibinfo{author}{\bibfnamefont{O.}~\bibnamefont{Painter}}, \bibnamefont{and}
  \bibinfo{author}{\bibfnamefont{H.~J.} \bibnamefont{Kimble}},
  \bibinfo{journal}{Phys. Rev. Lett.} \textbf{\bibinfo{volume}{108}},
  \bibinfo{pages}{214302} (\bibinfo{year}{2012}),
  \urlprefix\url{http://link.aps.org/doi/10.1103/PhysRevLett.108.214302}.

\bibitem[{\citenamefont{Thompson et~al.}(2008)\citenamefont{Thompson, Zwickl,
  Jayich, Marquardt, Girvin, and Harris}}]{Thompson:2008fk}
\bibinfo{author}{\bibfnamefont{J.~D.} \bibnamefont{Thompson}},
  \bibinfo{author}{\bibfnamefont{B.~M.} \bibnamefont{Zwickl}},
  \bibinfo{author}{\bibfnamefont{A.~M.} \bibnamefont{Jayich}},
  \bibinfo{author}{\bibfnamefont{F.}~\bibnamefont{Marquardt}},
  \bibinfo{author}{\bibfnamefont{S.~M.} \bibnamefont{Girvin}},
  \bibnamefont{and} \bibinfo{author}{\bibfnamefont{J.~G.~E.}
  \bibnamefont{Harris}}, \bibinfo{journal}{Nature}
  \textbf{\bibinfo{volume}{452}}, \bibinfo{pages}{72} (\bibinfo{year}{2008}).

\bibitem[{\citenamefont{Buchler et~al.}(1999)\citenamefont{Buchler, Gray,
  Shaddock, Ralph, and McClelland}}]{Buchler:1999fk}
\bibinfo{author}{\bibfnamefont{B.~C.} \bibnamefont{Buchler}},
  \bibinfo{author}{\bibfnamefont{M.~B.} \bibnamefont{Gray}},
  \bibinfo{author}{\bibfnamefont{D.~A.} \bibnamefont{Shaddock}},
  \bibinfo{author}{\bibfnamefont{T.~C.} \bibnamefont{Ralph}}, \bibnamefont{and}
  \bibinfo{author}{\bibfnamefont{D.~E.} \bibnamefont{McClelland}},
  \bibinfo{journal}{Opt. Lett.} \textbf{\bibinfo{volume}{24}},
  \bibinfo{pages}{259} (\bibinfo{year}{1999}),
  \urlprefix\url{http://ol.osa.org/abstract.cfm?URI=ol-24-4-259}.

\bibitem[{\citenamefont{Marquardt et~al.}(2007)\citenamefont{Marquardt, Chen,
  Clerk, and Girvin}}]{Marquardt2007}
\bibinfo{author}{\bibfnamefont{F.}~\bibnamefont{Marquardt}},
  \bibinfo{author}{\bibfnamefont{J.~P.} \bibnamefont{Chen}},
  \bibinfo{author}{\bibfnamefont{A.~A.} \bibnamefont{Clerk}}, \bibnamefont{and}
  \bibinfo{author}{\bibfnamefont{S.~M.} \bibnamefont{Girvin}},
  \bibinfo{journal}{Phys. Rev. Lett.} \textbf{\bibinfo{volume}{99}},
  \bibinfo{pages}{093902} (\bibinfo{year}{2007}),
  \urlprefix\url{http://link.aps.org/doi/10.1103/PhysRevLett.99.093902}.

\bibitem[{\citenamefont{Eberle et~al.}(2010)\citenamefont{Eberle, Steinlechner,
  Bauchrowitz, H\"andchen, Vahlbruch, Mehmet, M\"uller-Ebhardt, and
  Schnabel}}]{Eberle:2010fk}
\bibinfo{author}{\bibfnamefont{T.}~\bibnamefont{Eberle}},
  \bibinfo{author}{\bibfnamefont{S.}~\bibnamefont{Steinlechner}},
  \bibinfo{author}{\bibfnamefont{J.}~\bibnamefont{Bauchrowitz}},
  \bibinfo{author}{\bibfnamefont{V.}~\bibnamefont{H\"andchen}},
  \bibinfo{author}{\bibfnamefont{H.}~\bibnamefont{Vahlbruch}},
  \bibinfo{author}{\bibfnamefont{M.}~\bibnamefont{Mehmet}},
  \bibinfo{author}{\bibfnamefont{H.}~\bibnamefont{M\"uller-Ebhardt}},
  \bibnamefont{and} \bibinfo{author}{\bibfnamefont{R.}~\bibnamefont{Schnabel}},
  \bibinfo{journal}{Phys. Rev. Lett.} \textbf{\bibinfo{volume}{104}},
  \bibinfo{pages}{251102} (\bibinfo{year}{2010}),
  \urlprefix\url{http://link.aps.org/doi/10.1103/PhysRevLett.104.251102}.

\bibitem[{\citenamefont{Cole et~al.}(2013)\citenamefont{Cole, Zhang, Martin,
  Ye, and Aspelmeyer}}]{Cole:2013kx}
\bibinfo{author}{\bibfnamefont{G.~D.} \bibnamefont{Cole}},
  \bibinfo{author}{\bibfnamefont{W.}~\bibnamefont{Zhang}},
  \bibinfo{author}{\bibfnamefont{M.~J.} \bibnamefont{Martin}},
  \bibinfo{author}{\bibfnamefont{J.}~\bibnamefont{Ye}}, \bibnamefont{and}
  \bibinfo{author}{\bibfnamefont{M.}~\bibnamefont{Aspelmeyer}},
  \bibinfo{journal}{Pre-print} \textbf{\bibinfo{volume}{arXiv:1302.6489}}
  (\bibinfo{year}{2013}).

\bibitem[{\citenamefont{Miao et~al.}(2010)\citenamefont{Miao, Danilishin,
  M{\"u}ller-Ebhardt, and Chen}}]{Miao:2010zr}
\bibinfo{author}{\bibfnamefont{H.}~\bibnamefont{Miao}},
  \bibinfo{author}{\bibfnamefont{S.}~\bibnamefont{Danilishin}},
  \bibinfo{author}{\bibfnamefont{H.}~\bibnamefont{M{\"u}ller-Ebhardt}},
  \bibnamefont{and} \bibinfo{author}{\bibfnamefont{Y.}~\bibnamefont{Chen}},
  \bibinfo{journal}{New Journal of Physics} \textbf{\bibinfo{volume}{12}},
  \bibinfo{pages}{083032} (\bibinfo{year}{2010}),
  \urlprefix\url{http://stacks.iop.org/1367-2630/12/i=8/a=083032}.

\bibitem[{\citenamefont{Wilson-Rae et~al.}(2007)\citenamefont{Wilson-Rae,
  Nooshi, Zwerger, and Kippenberg}}]{Wilson-Rae2007}
\bibinfo{author}{\bibfnamefont{I.}~\bibnamefont{Wilson-Rae}},
  \bibinfo{author}{\bibfnamefont{N.}~\bibnamefont{Nooshi}},
  \bibinfo{author}{\bibfnamefont{W.}~\bibnamefont{Zwerger}}, \bibnamefont{and}
  \bibinfo{author}{\bibfnamefont{T.~J.} \bibnamefont{Kippenberg}},
  \bibinfo{journal}{Phys. Rev. Lett.} \textbf{\bibinfo{volume}{99}},
  \bibinfo{pages}{093901} (\bibinfo{year}{2007}),
  \urlprefix\url{http://link.aps.org/doi/10.1103/PhysRevLett.99.093901}.

\bibitem[{\citenamefont{Genes et~al.}(2008)\citenamefont{Genes, Vitali,
  Tombesi, Gigan, and Aspelmeyer}}]{Genes2008a}
\bibinfo{author}{\bibfnamefont{C.}~\bibnamefont{Genes}},
  \bibinfo{author}{\bibfnamefont{D.}~\bibnamefont{Vitali}},
  \bibinfo{author}{\bibfnamefont{P.}~\bibnamefont{Tombesi}},
  \bibinfo{author}{\bibfnamefont{S.}~\bibnamefont{Gigan}}, \bibnamefont{and}
  \bibinfo{author}{\bibfnamefont{M.}~\bibnamefont{Aspelmeyer}},
  \bibinfo{journal}{Phys. Rev. A} \textbf{\bibinfo{volume}{77}},
  \bibinfo{pages}{033804} (\bibinfo{year}{2008}),
  \urlprefix\url{http://link.aps.org/doi/10.1103/PhysRevA.77.033804}.

\bibitem[{\citenamefont{Milburn and Woolley}(2011)}]{Milburn2011}
\bibinfo{author}{\bibfnamefont{G.~J.} \bibnamefont{Milburn}} \bibnamefont{and}
  \bibinfo{author}{\bibfnamefont{M.~J.} \bibnamefont{Woolley}},
  \bibinfo{journal}{Acta Physica Slovaca} \textbf{\bibinfo{volume}{61}},
  \bibinfo{pages}{483} (\bibinfo{year}{2011}).

\bibitem[{\citenamefont{Walls and Milburn}(2008)}]{Walls2008}
\bibinfo{author}{\bibfnamefont{D.~F.} \bibnamefont{Walls}} \bibnamefont{and}
  \bibinfo{author}{\bibfnamefont{G.}~\bibnamefont{Milburn}},
  \emph{\bibinfo{title}{Quantum Optics}} (\bibinfo{publisher}{Springer-Verlag},
  \bibinfo{year}{2008}).

\bibitem[{\citenamefont{Drever et~al.}(1983)\citenamefont{Drever, Hall,
  Kowalski, Hough, Ford, Munley, and Ward}}]{Drever:1983fk}
\bibinfo{author}{\bibfnamefont{R.~W.~P.} \bibnamefont{Drever}},
  \bibinfo{author}{\bibfnamefont{J.~L.} \bibnamefont{Hall}},
  \bibinfo{author}{\bibfnamefont{F.~V.} \bibnamefont{Kowalski}},
  \bibinfo{author}{\bibfnamefont{J.}~\bibnamefont{Hough}},
  \bibinfo{author}{\bibfnamefont{G.~M.} \bibnamefont{Ford}},
  \bibinfo{author}{\bibfnamefont{A.~J.} \bibnamefont{Munley}},
  \bibnamefont{and} \bibinfo{author}{\bibfnamefont{H.}~\bibnamefont{Ward}},
  \bibinfo{journal}{Appl. Phys. B} \textbf{\bibinfo{volume}{31}},
  \bibinfo{pages}{97} (\bibinfo{year}{1983}).

\end{thebibliography}

\end{document}